\documentclass[nohyper,11pt,a4paper]{JHEP3}
\pdfoutput=1
\usepackage{graphicx}
\usepackage{bm}
\usepackage{mathrsfs}
\usepackage{amsmath}
\usepackage{amssymb}
\usepackage{amstext}
\usepackage{epsfig,latexsym}
\newcommand{\DFIGURE}[1]{\begin{figure}[htbp]#1\end{figure}}
\newcommand{\MFIGURE}[3]{\FIGURE{\includegraphics[width=0.49\textwidth]{#1}\caption{#2}\label{#3}}}

\newcommand{\cQ}{q_\mt{7}} 

\newcommand{\be}{\begin{equation}}
\newcommand{\ee}{\end{equation}}
\newcommand{\bea}{\begin{eqnarray}}
\newcommand{\eea}{\end{eqnarray}}
\newcommand{\ba}{\begin{eqnarray}}
\newcommand{\ea}{\end{eqnarray}}

\newcommand{\beq}{\begin{equation}}
\newcommand{\eeq}{\end{equation}}
\newcommand{\beqa}{\begin{eqnarray}}
\newcommand{\eeqa}{\end{eqnarray}}
\newcommand{\beqar}{\begin{eqnarray*}}
\newcommand{\eeqar}{\end{eqnarray*}}

\newcommand{\ie}{{\it i.e.,}\ }

\newcommand{\tlt}{{\tilde{t}}} 
\newcommand{\tx}{{\tilde{x}}}

\newcommand{\tlc}{{\tilde{c}}}
\newcommand{\tlm}{{\tilde{m}}}
\newcommand{\tlb}{{\tilde{B}}}
\newcommand{\tlrho}{{\tilde{\rho}}}

\def\nc {N_\mt{c}}
\def\nf {N_\mt{f}}

\def\t6 {T_\mt{D6}}

\newcommand{\ls}{\ell_\mt{s}}

\newcommand{\mt}[1]{\textrm{\tiny #1}}

\newcommand{\bi}{\begin{itemize}}
\newcommand{\ei}{\end{itemize}}
\newcommand{\ben}{\begin{enumerate}}
\newcommand{\een}{\end{enumerate}}

\newcommand{\bmtx}{\left[ \begin{array}{cc}}
\newcommand{\emtx}{\end{array} \right]}
\newcommand{\bvec}{\left[ \begin{array}{c}}
\newcommand{\evec}{\end{array} \right]}

\newcommand{\bfig}{\begin{figure}}
\newcommand{\efig}{\end{figure}}

\newcommand{\order}{{\mathcal{O}}}

\newcommand{\labell}[1]{\label{#1}}
\newcommand{\reef}[1]{(\ref{#1})}
\newcommand{\mref}[1]{\ref{#1}}
\newcommand{\mlabel}[1]{\label{#1}}
\title{Thermodynamics of Holographic Defects}
\author{
Matthias C. Wapler\\
Perimeter Institute for Theoretical Physics,
Waterloo, Ontario N2L 2Y5, Canada \\
Department of Physics \& Astronomy and Guelph-Waterloo Physics
Institute,\\
\ \ \ University of Waterloo,
Waterloo, Ontario N2L 3G1, Canada \\
Center for Quantum Spacetime, Sogang University, Seoul, Korea \\
\\E-mail: \email{wapler@sogang.ac.kr}}
\preprint{}
\date{\today}
\abstract{%
Using the AdS/CFT correspondence, we study the thermodynamic properties and the phase diagram of matter fields on (2+1)-dimensional defects coupled to a (3+1)-dimensional ${\cal N}=4$ SYM ``heat bath''. 
Considering a background magnetic field, (net) quark density, defect ``magnitude'' $\delta N_c$ and the mass of the matter, we study the defect contribution to the thermodynamic potentials and their first and second derivatives to map the phases and study their physical properties.

We find some features that are qualitatively similar to other systems e.g. in (3+1) dimensions and
a number of features that are particular to the defect nature, such as its magnetic properties, unexpected properties at $T\rightarrow 0$ and finite density; and the finite $\delta N_c$ effects, e.g. a diverging susceptibility and vanishing density of states at small temperatures, a
physically consistent negative heat capacity and new types of consistent phases.}
\begin{document}{\vskip 1cm}
\section{Introduction}
In recent years, there has been a great amount of activity related to applying the AdS/CFT correspondence \cite{juan,adscft,bigRev} to scenarios that may have some relevance in the context of experimental physics.
While the first and most common example, the duality between a stack of $N_c$ D3 branes generating an $AdS_5 \times S^5$ geometry in the decoupling limit and a thermal $\mathcal{N} = 4$ $SU(N_c)$ super-Yang-Mills theory on its boundary may be of limited experimental relevance,
for example  fields transforming in the fundamental representation of the $SU(N_c)$ have been studies by introducing a small number of $N_f$ ``probe'' D7-branes into the background, covering all of the AdS directions \cite{lisa,karchkatz}.
Certainly this setup is still significantly different from QCD, but this model and the T-dual D4-D6 setup have received great interest \cite{recent,johanna}, in particular the thermodynamics and phase structure \cite{long} and the ``meson spectrum'' \cite{meson} -- hoping that some results obtained from AdS/CFT may be sufficiently generic, such that they also apply to QCD.

More recently there has also been great effort on applying the AdS/CFT correspondence to condensed matter physics which may be more promising, since there is only one QCD but there are on the one hand many different strongly coupled effective field theories in condensed matter physics and on the other hand there exist a large number AdS string vacua \cite{denef}. 
The vast majority of these efforts have been related to 2+1 dimensional field theories.
The first example was an $AdS_4 \times S^6$ geometry obtained from an M2-brane setup in \cite{pavel} to study some transport properties. Since then, many interesting properties have been studied such as Hall conductivity \cite{hallo,mmore}, superconductivity and superfluidity \cite{more, magsup} or the Nernst effect \cite{nernst}, mostly in setups that are not based on string theory and hence do not ensure that the systems are pathology-free. Another approach to move towards experimentally relevant field theories has been the construction of duals of non-relativistic CFTs \cite{nonrel}. A not yet satisfactorily addressed question is what the implications of the Fermi-Dirac distribution is in AdS/CFT \cite{fermi}.

The most common example of how systems with conformal symmetry arise in condensed matter physics is the quantum critical phase. This phase arises in the context of a phase transition at zero temperature, a so-called ``quantum critical phase transition'' at a ``quantum critical point''. At this point, the system displays scale-invariant behavior as also in other phase transitions, but in contrast to phase transitions at finite temperature, it extends into a whole region in the phase diagram that may be described by a conformal field theory, the so-called quantum critical phase; shown in fig. \ref{surfdi}. 
\cite{subir,sachdev}.

All of the above-mentioned applications of the AdS/CFT correspondence to condensed matter physics have in common that they are described by 2+1 dimensional field theories. In our 3+1 dimensional world however, all 2+1 dimensional systems are strictly speaking defects. In some cases this fact may be less relevant and in other cases more relevant. Hence, it is interesting to study the physics of a 2+1 dimensional defect in order to explore what difference there is to purely 2+1 dimensional systems. Defect field theories are basically field theories in which matter that is confined to some hypersurface interacts via a field theory in the bulk. 
While there is some review literature in a soft condensed matter context
(for a review see \cite{soft}) there seems to be not much review literature related to the defects and their aspects that we are interested in.
Hence, a motivated guess may be that they have many properties in common with surfaces, which have been studied extensively. 
To illustrate their properties, we can look in fig. \mref{surfdi} at a generic surface phase diagram of some system described by a bulk coupling $J_b$ and a  surface coupling $J_s$ -- for a review on the subject see \cite{surftrans}. There we see that over most of the parameter range the surface and the bulk are in the same phase and display a simultaneous ``ordinary'' phase transition. As we tune the surface coupling beyond a ``special point'', which is some critical multiple of the bulk coupling, the phase transitions on the surface and in the bulk separate into a surface phase transition and an ``extraordinary'' phase transition in which there is a phase transition only in the bulk. It is obvious from the ratio $J_s/J_b$ in this regime that the ordered phase on the surface extends to higher temperatures than the ordered phase in the bulk. However it is quite interesting that this splitting of phase transitions typically occurs as $J_s$ becomes greater than $J_b$ and hence there is no ``mirror symmetric'' version of this plot.

\DFIGURE{
\includegraphics[width=0.49\textwidth]{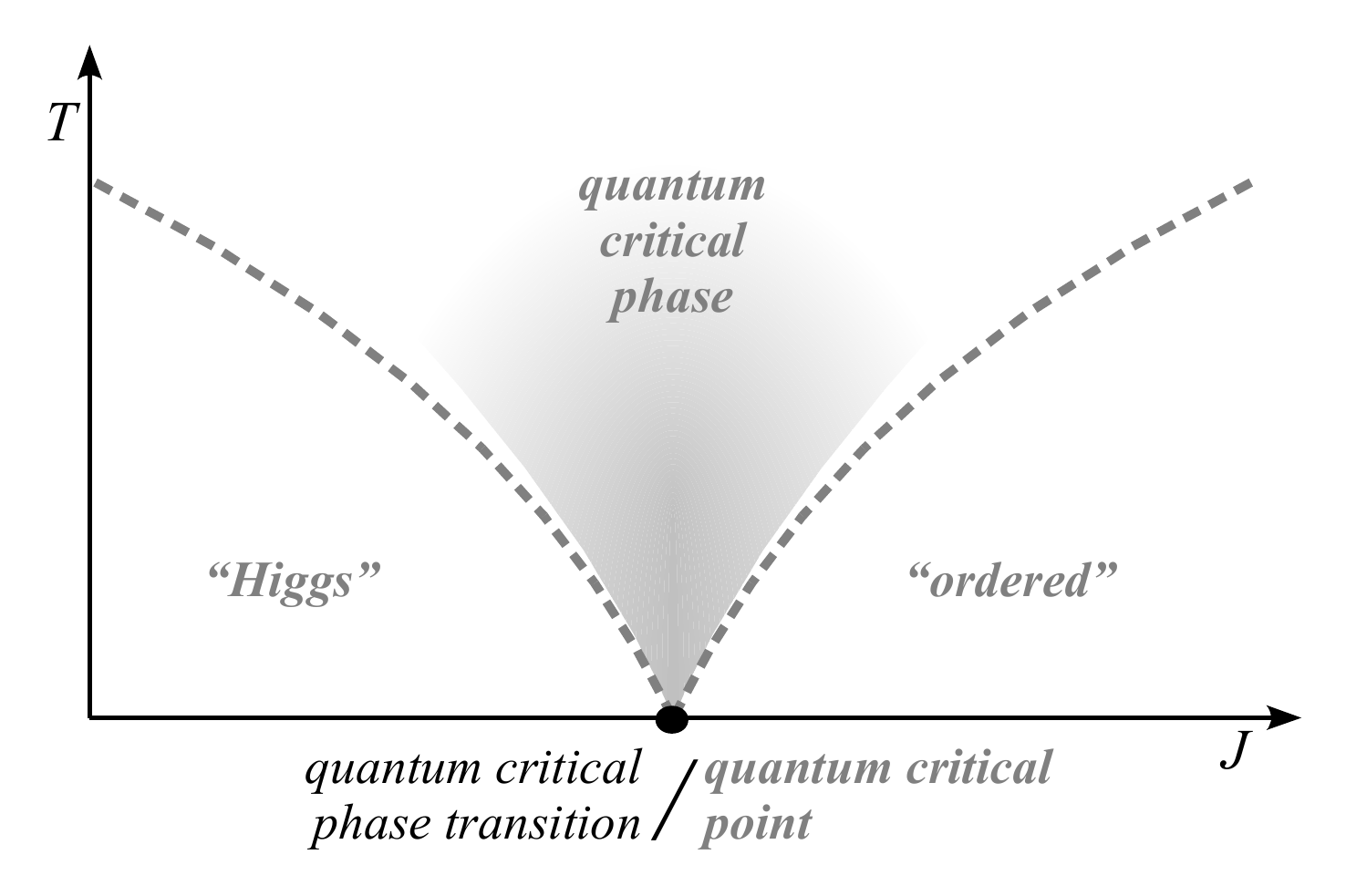}
\includegraphics[width=0.49\textwidth]{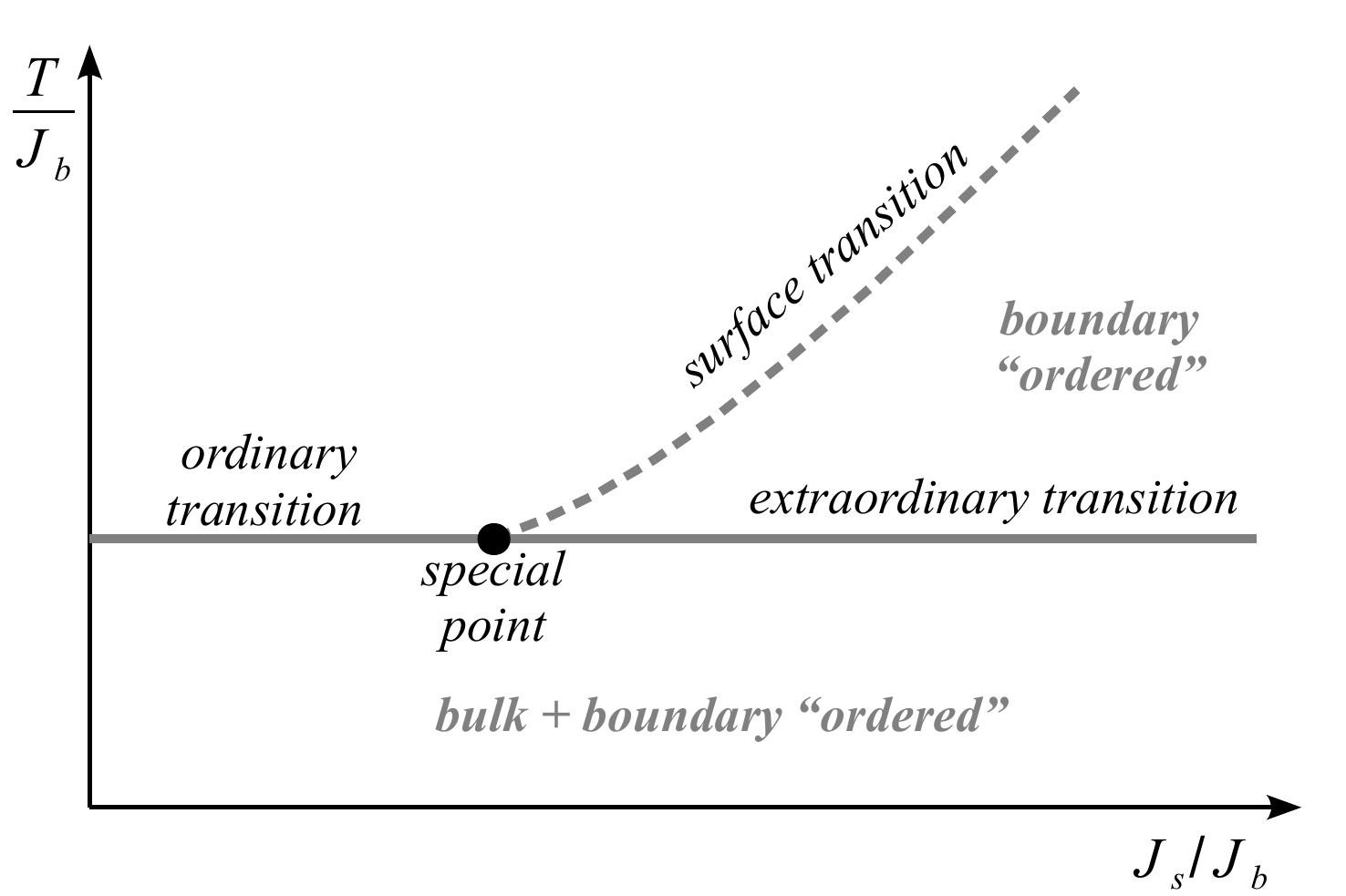}
\caption{Left: A generic quantum critical phase transition and quantum critical phase. Right: A generic surface phase diagram.}\mlabel{surfdi}}
In this paper, we study the thermodynamics and phase diagram of the defect.
This means, we have to consider the contributions to the thermodynamic and statistical quantities of the full (defect+bulk) field theory that are extrinsic in the area of the defect.
Certainly, we will miss contributions that are extrinsic in the volume of the SYM, but these depend only on a ``topological'' parameter that is related to a difference in the level of the gauge group $N_c \rightarrow \delta N_c$ between both sides of the defect. As indicated in \cite{baredef}, this parameter is also related to a ``width'' of the region around the defect. There it was observed from the quasiparticle spectrum, and the length scales found in various regimes, whereas here we will discuss it from a more direct point of view. It will also be relevant when we argue about the positivity of the entropy and the heat capacity. As in \cite{fancycon}, the other parameters are the net baryon number density $\rho$, an externally applied magnetic field $B$ and the ``quark'' mass $M_q$.

To study the thermodynamics, we obtain the free energy from the euclidean action, from which we obtain the chemical potentials and response functions. We also demonstrate how to obtain the second derivatives of the thermodynamic potential in closed form, i.e. without having to use numerical derivatives; even in the case of finite $M_q$.

Using those quantities, we will map out the phase diagram and point out some important differences arising from the choice of the ``free'' thermodynamic variables. We then argue that the blackhole phase studied in \cite{fancycon} is indeed the thermodynamically preferred one and study the physical properties of the different phases. We will then also find in some region of phase space a first-order transition between two distinct regions of the blackhole phase. It will turn out that this is in some sense a smooth continuation of the blackhole-Minkowski embedding phase transition, that was extensively studied in 3+1 dimensional systems and beyond \cite{long,johanna,findens,paolo}. In contrast to the observations in those systems, it will turn out that a finite density Minkowski embedding can be realized in our case also at finite density, even though it is only metastable. Above the phase transition, the physics that we find will be dominated by some simple scaling laws, but below, in the small-temperature regime, we will find some surprising non-trivial effects.

The gravitational setup of the defect is a stack of $N_f$ probe D5- or D7-branes inserted into the background of a stack of $N_c$ D3-branes and is well known from the literature \cite{lisa,hirosi,jaume,rey,quantumhall}.
As in \cite{baredef,fancycon}, the difference $\delta N_c$ in the level of the gauge group of the 3+1 SYM will be introduced by an additional flux on the probe brane in the compact sphere, which also stabilizes the D7 setup.
Similarly, the finite magnetic field and net density are introduced using the well-known duals of a magnetic field and an electric field, respectively, in the world volume of the probe brane.
The finite quark mass will be obtained by a deformation of the embedding in the compact sphere in the same fashion in which it was done in the duals for 3+1 dimensional QCD-like systems \cite{long,findens,us,johanna}. 

The (3+1)-dimensional system of the ${\cal N}=2$ gauge theory
constructed with parallel D7- and D3-branes (see
e.g. \cite{johanna,recent,long}) is a generically good reference to which to compare our analysis. There, it was found that if a finite $M_q$ is introduced, the scale
$M_\mt{fun}\sim M_q/\sqrt{\lambda}$ plays a special role in this
theory. At vanishing density, the picture is as follows: First, the ``mesons", bound states of a fundamental and an
anti-fundamental field, are deeply bound with their spectrum of
masses characterized by $M_\mt{fun}$ \cite{meson}. Next at a
temperature $T\sim M_\mt{fun}$, the system undergoes a phase
transition characterized by the dissociation of the mesonic bound
states \cite{long,johanna}. 
In the presence of background magnetic fields, it was found that the phase transition moves to larger Temperatures \cite{magnetic}. Some aspects of the (2+1) dimensional conformal matter in the presence of a magnetic field have also been studied in \cite{Filev}, which appeared on the arXiv during the closing stages of this work.

The outline of this paper is as follows:
In section \mref{gravsetup}, we briefly review the construction of the defect setup and the AdS/CFT dictionary for the background quantities and point out several problems that arise in the D3-D7 setup -- which motivate us not to pursue the massive D7 case. 
We then discuss the thermodynamics in section \ref{thermdyn}, starting with a discussion of the thermodynamic potential and the thermodynamic variables in section \ref{thermdefs}, from which we then obtain the response functions in section \ref{responses}, which we discuss first in the massless case in \mref{massless} and then in the massive case in \mref{massive}. In the latter case, we first outline the phase diagram in 
 section \mref{phases} and then discuss the physics of the response functions in \mref{pots}. 
\section{Gravitational setup}\mlabel{gravsetup}
In this section, we briefly review the string theory setup of the defect that was outlined in detail in \cite{fancycon}. We also point out some minor differences in the case of non-blackhole embeddings.
\subsection{$D3$ = (N=4 SYM) background}\mlabel{fancysetup}
The well-known $AdS_5 \times S^5$ background from the decoupling limit of $N_c$ D3 branes at finite temperature $T = \frac{r_0}{\pi L^2}$ can be written as 
\begin{equation}
ds^2 \, = \, \frac{r^2}{L^2}\left(- h(r) dt^2 + d\vec{x}_3^2  \right) + \frac{L^2}{r^2}\left(\frac{dr^2}{h(r)} + r^2 d\Omega_5^2 \right)\ , ~~~~~ C^{(4)}_{\!\! 0123}= -\frac{r^4}{L^4} \, 
\end{equation}
and corresponds to  an $N = 4$ SYM theory on the boundary with $U(N_c)$ gauge group  \cite{juan,adscft,bigRev}. We work, as usual, in the limit $N_c \rightarrow \infty$, Yang-Mills coupling $g_{YM}^2 = 2\pi g_s \rightarrow 0$ and t'Hoft coupling $\lambda = g^2_{YM} N_c \to \infty$, i.e. we are in the supergravity limit $L^4 = 4\pi g_s N_c l_s^4 \rightarrow \infty$.
Considering only $T > 0$ allows us to go to dimensionless coordinates $u = \frac{r_0}{r}, \ \tlt = \frac{ r_0 t }{L^2}, \ \vec{\tx} = \vec{x}\frac{ r_0}{L^2}$:
\begin{equation}
ds^2 \, = \, \frac{L^2}{u^2} \left(- (1-u^4)d\tlt^2 + d\vec{\tx}_3^2 + \frac{du^2}{1 - u^4} + u^2 d\Omega_5^2 \right) .
\end{equation}

The fields in the SYM all transform in the adjoint representation of the $SU(N_c)$, and we follow the well-known probe brane approach of introducing $N_f$ families of fields transforming in the fundamental representation, in the ``quenched approximation'' $N_f \ll N_c$. Commonly, one inserts for example $N_f$ D7 branes in the D3 background parallel to the D3's \cite{karchkatz}. Generalizations to defect configurations have been considered in the literature \cite{lisa,hirosi,jaume}.
In the latter cases,
the intersection overs only part of the flat directions, 
creating the
defect field theory, where the fundamental fields are
only supported on a subspace within the four-dimensional spacetime
of the gauge theory. 
In the case of our
$(2+1)$-dimensional defect, this can be done with the following configuration,
\begin{equation}
\begin{array}{rccccc|c|cccccl}
  & & 0 & 1 & 2 & 3 & 4& 5 & 6 & 7 & 8 & 9 &\\
  & & t & x & y & z & r&   &   &   &   &  \psi &\\
\mathrm{background\,:}& D3 & \times & \times & \times & \times & & &  & & & & \\
\mathrm{probe\,:}& D5 & \times & \times & \times &  & \times  & \times & \times & &  & &  \\
& D7 & \times & \times & \times &  & \times  & \times & \times & \times & \times & &  \ \ \ . \\
\end{array} 
\labell{array}
\end{equation}

The D5-brane construction reduces the supersymmetry from ${\cal N}=4$ of the D3 background to
${\cal N}=2$ and the dual field
theory is now the SYM gauge theory coupled to $\nf$ fundamental
hypermultiplets, which are confined to a (2+1)-dimensional defect.
In the D7
case, all supersymmetry is broken and the defect CFT contains only
$\nf$ flavors of fermions, again in the fundamental representation
\cite{rey}. 
It turns out that the lack of supersymmetry in the
latter case will manifest itself with the appearance of
instabilities. In \cite{baredef}, we showed how this can be avoided, and we discussed in \cite{fancycon}  how this instability becomes apparent in the scaling dimension of the scalar field that corresponds to the deformation of the $S^4$ of the D7-worldvolume inside the $S^5$ background. There, we also discussed some problems related to the reliability of the quenched approximation that we consider in this paper.

In this limit, $\nf
\ll \nc$, the D5-branes may be treated as probes in the
supergravity background, \ie we may ignore the gravitational
back-reaction of the branes. For the D7-branes, however, this is only true locally and not in the asymptotic regime.

\subsection{Introducing the defect}\mlabel{fancyintro}
Considering only the $U(1)$ subgroup of the $U(N_f)$, the action of the D5 brane in a D3 background is just the DBI action plus a Chern-Simons term 
\begin{equation}
S \, = \, - T_5 N_f \int_{D5} \sqrt{- det(P[G] + 2\pi l_s^2 F )} \, +\, T_5 N_f \int_{D5} C^{(4)} \wedge 2\pi l_s^2 F  \ ,
\end{equation}
where the factors of $N_f$ arise from taking the trace over the flavor degrees of freedom.

Assuming  translational (in the flat directions) and rotational (on the sphere) symmetry, the induced metric of the embedding \reef{array} can be written in the form
\begin{eqnarray}\labell{branemet}
ds^2 &=& \frac{L^2}{u^2} \left( -(1- u^4)d\tlt^2  + d\vec{\tx}_2^2  + \left(1+(1-u^4)\left(z\chi'(u)^2   + u^2\frac{\Psi'(u)^2}{1-\Psi(u)^2} \right) \right) \frac{du^2}{1-u^4}  \nonumber \right. \\ & & \left. \ \ \ \ \ \ \ \ \ \ \, + \, u^2 (1-\Psi(u)^2) d\Omega_2^2\right) \ ,
\end{eqnarray} 
where we used the inverse radius $u = \frac{r_0}{r}$ and $\chi = \frac{z \,r_0}{L^2} = \pi T\, z$.
Here and in the rest of the paper, we also use the notation $\tilde{\cdot}$ to denote quantities that are made dimensionless with appropriate factors of $\pi T$ unless explicitly noted otherwise.

Now, we wish to consider a few interesting background quantities.
Firstly, we can turn on a magnetic flux on the sphere, $F=\frac{q}{N_f} d\Omega_2$, which corresponds to inducing an extra number of colors, $\delta N_c = q \in \mathbb{Z}$, parametrized by the quasi-continuous $f :=\frac{\pi l_s^2}{L^2 N_f} q = \frac{\pi q}{\sqrt{\lambda} N_f}$, on one side of the defect and causes the probe brane to bend towards that side. Both from the embedding geometry $\partial_u \chi(u)=\frac{- f}{\sqrt{1+f^2 u^4}}$, and from the resulting quasiparticle spectrum in the field theory, it was argued in \cite{baredef} that this flux also introduced a finite width, $\Delta z$, of the defect. 

Using the AdS/CFT dictionary in \cite{bigRev} in analogy with the $3+1$ dimensional system, e.g. \cite{findens,magnetic,hallo}, we find 
the gravity dual of the baryon density
\begin{equation}
\rho  \ = \ - \frac{\delta S}{\delta A_{0}^{bdy.}} \ = \  \frac{\sqrt{\lambda}N_c T^2}{4 \pi } \lim_{u\rightarrow 0} \partial_u \tilde{A}_0(u)
\end{equation}
and magnetic field $B$ to be related to a non-trivial $U(1)$ background on the brane:
\begin{equation}\labell{eback}
F|_{u \rightarrow 0} \, = \, - E(u) dt\wedge dr  + B dx\wedge dy \, =: \, F_E + F_B .
\end{equation}
We can also define the (asymmetric) background metric
\beq \labell{gdef}
G = g+ F \ .
\eeq

A finite quark mass $M_q$ and dual condensate $C$ can be associated with a non-trivial embedding $\Psi(u)$
by analogy with the $3+1$ dimensional $D3-D7$ system \cite{long,findens,us,johanna}.
This condensate has on the one hand an interpretation as a chemical potential for $M_q$ and on the other hand is considered in QCD contexts as the order parameter of chiral symmetry breaking.
Parametrizing the $S^5$ as $d\Omega_5 ^2 = d\psi ^2 + \cos^2 \psi
\, d\Omega_2^2 \,+\sin^2 \psi \, d{\Omega}_2^2$ and putting the D5 on the first $S^2$, such that $\sin \psi  =:  \Psi$, the DBI-CS action becomes
\begin{eqnarray}\labell{backact}
S &=& 4\pi L^2  T_5  \int\! d\sigma^4 \left(\sqrt{-\det G}\sqrt{(1-\Psi^2)^2 + f^2} + f u^4 \partial_u z \right) \\ \nonumber &=& 4 \pi L^2 T_5 \int d\sigma^4\left( \sqrt{-\det G}\sqrt{1+F_{E}^2}\sqrt{1+F_{\tlb}^2}\sqrt{(1-\Psi^2)^2 + f^2}  + f u^4 \partial_u z\right).
\end{eqnarray}
The background solution is then
\begin{eqnarray}\labell{backgd}
B &=& const. \\ 
\partial_u \tilde{A}_0(u) &=&  \frac{\tlrho \sqrt{1+f^2}\sqrt{1- \Psi^2(u) + u^2 h(u) \Psi'(u)^2}}{\sqrt{1-\Psi(u)^2}\sqrt{1+\big(  f^2 +(\tlrho^2+ \tlb^2)(1+f^2)\big)u^4 + (1+\tlb^2 u^4)\Psi(u)^2(\Psi(u)^2-2)}} \nonumber \\
\partial_u \chi & = &  \frac{- f \sqrt{1- \Psi^2(u) + u^2 h(u) \Psi'(u)^2}}{\pi T \sqrt{1-\Psi(u)^2}\sqrt{1+\big( f^2+(\tlrho^2 + \tlb^2)(1+f^2)\big)u^4 + (1+\tlb^2 u^4)\Psi(u)^2(\Psi(u)^2-2)}} \nonumber,
\end{eqnarray} 
where $h(u) = 1-u^4$, $\tlrho :=   \frac{4 \pi \rho}{\sqrt{\lambda}N_c N_f T^2}$ and $\tlb = \frac{B}{(\pi T)^2}$.
Now, the width of the defect from the brane picture, $z_{max} := \lim_{u->1} z(u)$, decreases with increasing density or magnetic field, in agreement with the quasiparticle spectrum observed in \cite{fancycon}.

The equation of motion for $\Psi(u)$ becomes
\begin{eqnarray}\labell{psieom}
\!\!\!\!\!\!\!\!\!\!\frac{2(1+\tlb^2u^4)(1-\Psi^2)^3 +u^2 (1-u^4)\big(1-(f^2 + (\tlrho^2+\tlb^2)(1+f^2))u^4 + (1+\tlb^2 u^4)\Psi^2(\Psi^2-2) \big)\Psi'^2}{u^4 (1\!-\!\Psi^2)\sqrt{(1\!-\!\Psi^2)\big(1\!-\!\Psi^2+(u^2\!-u^6)\Psi'^2 \big)\big(1+( f^2\!+(\tlrho^2\! + \tlb^2)(1+f^2))u^4 + (1+\tlb^2 u^4)\Psi^2(\Psi^2\!-2)\big)}} \nonumber \\
\!\!\!\!\!\!\!\!\!\! \!\!\!\!\!\!\!\!\!\! = \
\partial_u  \left(\Psi' \frac{1-u^4}{u^2}\sqrt{\frac{1+( f^2+(\tlrho^2 + \tlb^2)(1+f^2))u^4 + (1+\tlb^2 u^4)\Psi^2(\Psi^2-2)}{(1-\Psi^2)(1-\Psi^2+(u^2-u^6)\Psi'^2)}} \right)~~~~~~~~~~~~~
 \ , 
\end{eqnarray}
which has no analytical solution, except for some limiting cases. For $u \rightarrow 0$, it is easy to see that the solution becomes
\begin{equation}\labell{psiasym}
\Psi \, \sim \, \tlm \, u \, + \, \tlc \, u^2 \ ,
\end{equation}
where $\tlm$ and $\tlc$ are dimensionless free parameters that are determined by the boundary conditions at $u=1$. Following arguments of the T-dual $(3+1)$ dimensional D3-D7 \cite{long,findens,us,johanna}, the quark mass $M_q$ and condensate $C$ are given by 
\begin{equation}
M_q \, = \, \frac{r_0 \,\tlm}{2^{3/2}\pi l_s^2} \, = \, \sqrt{\lambda} \frac{T}{2^{3/2}}  \tlm\ \ \  \mathrm{and}  \ \ \  C\, = \, \sqrt{2}4\pi^2 \, r_0^2 N_f l_s^2 T_5 \tlc \, = \,  \frac{1}{4\pi} T^2 N_f N_c\tlc\ .
\end{equation}
In section \ref{thermdefs}, we discuss this more in detail and verify that $C$ is indeed the dual chemical potential to the mass.

In order to find the solution for the full geometry for a given mass however, we need consider the equation near the horizon, where \reef{psieom} reduces to first order,
\begin{equation}
\left.\Psi'\right|_{u\rightarrow 1} \, =\, \frac{1}{2}\frac{(1+\tlb^2)\Psi_0 (1-\Psi_0^2)^2}{(1-\Psi_0^2)^2+f^2+\tlrho^2+\tlb^2\big(1 + (1-\Psi_0^2)^2\big)} \ ,
\end{equation}
effectively relating $\tlm$ and $\tlc$. Hence, the only remaining boundary condition at the horizon is  $\Psi_{u\rightarrow 1} = \Psi_0$. We then have to find recursively $\Psi_0$ for a given value of $\tlm$. 

So far, this embedding is the same as the the one discussed in \cite{fancycon}. Now, however, we also want to consider non-blackhole embeddings. In this case, the probe brane does not extend to the horizon $u=1$, but closes off, i.e. $\Psi \rightarrow 1 $, at some finite value of $u$, $u_{max}$, above the horizon. The above discussion carries over, but now we are tuning $u_{max}$ instead of $\Psi_0$. Just as at the horizon, the equations of motion reduce to first order at this point, and the boundary conditions at $u_{max}$ can be read off from the expansion 
\begin{equation}\labell{umax}
\Psi \  = \ 1 - \frac{u_{max} -u}{u_{max}(1+u_{max}^{\! 4})} + \frac{(u_{max} -u)^2 u_{max}^{\! 2}(3+13 u_{max}^{\! 4})}{6 (1-u_{max}^{\! 4})(1+u_{max}^{\! 4})^2} + \order (u_{max} -u)^3 \ .
\end{equation} 
While in the case of the black hole embedding, the boundary condition at $u=1$ and the equations of motion ensure that $\Psi \in [0,1[$ over the whole range of $u$, we now have to discard unphysical embeddings in which $\Psi < 0$ at some values of $u$. These solutions appear at large magnetic fields and are in practice related to having $\left. \Psi'\right|_{u\rightarrow 0} < 0$, i.e. they would correspond to a negative value of the mass parameter.

In the non-supersymmetric case, the magnetic flux on the compact $S^2$ is replaced by an instanton on the compact $S^4$, and the coupling to the five-form flux comes now via the Chern-Simons term
\begin{equation}
\frac{(2\pi l_s^2)^2}{2}T_7 N_f \int_{D7} C^{(4)} \wedge  F \wedge  F \ .
\end{equation}
We also parametrize the $S^5$ slightly differently as $d\Omega_5 ^2 = d\psi ^2 + \cos^2 \psi \, d\Omega_4^2$, such that the induced metric is 
\begin{eqnarray}\labell{branemettt}
ds^2 &=& \frac{L^2}{u^2} \left( -(1- u^4)d\tlt^2  + d\vec{\tx}_2^2  + \left(1+(1-u^4)\left(z'(u)^2   + u^2\frac{\Psi'(u)^2}{1-\Psi(u)^2} \right) \right) \frac{du^2}{1-u^4}  \nonumber \right. \\ & & \left. \ \ \ \ \ \ \ \ \ \ \, + \, u^2 (1-\Psi(u)^2) d\Omega_4^2\right) \ .
\end{eqnarray}
It turns out that a qualitative difference arises only in the massive case, as in that case the five-form pulls back to the $S^4$ factor of the brane as 
 \begin{equation}
\frac{(2\pi l_s^2)^2}{2}T_7 N_f \int_{D7} F^{(5)} \wedge  A \wedge  F  \ \ \rightarrow \ \ 
8 T_7 N_f \frac{\pi^5 l_s^4}{L^4} \int_0^1 du \frac{\Psi'}{\sqrt{1-\Psi^2}}  \int_{\mathbb{R}^{(2,1)}} A \wedge  F  \ .
 \end{equation}
This was used in \cite{quantumhall} to obtain a Hall effect in a setup that would be considered from the perspective of this paper unstable. 

It is a straightforward exercise to verify that it turns out that in the massless case, the embedding and the solutions for background fields take precisely the same form as the ones in the supersymmetric case, under an appropriate re-definition of the flux parameter $f_7 \equiv \frac{Q}{\sqrt{1+ 2|Q|}}$ in terms of the instanton number $\frac{\lambda \nf Q}{6\pi^2}   =  \cQ = \frac{1}{8\pi^2}\oint_{S^4} Tr F\wedge F \in \mathbb{Z}$. Furthermore, even the value of the action, which reads now $\oint_{S^4} d^4\Omega  = \frac{8 \pi^2}{3}\left(\nf L^4 (1 - \Psi^2) + 6 \pi^2 \ls^4 |\cQ| \right)$, is identical, modulo an overall factor. 

A problem related to the stability of the D3-D7 solutions was pointed out in \cite{fancycon}. The ``smooth'' instanton solution discussed in \cite{neil2,ConstableRobEtc} that preserves the rotational symmetries of the $S^4$, and hence also the symmetries of our field theory, limits $\frac{N_f(N_f^2-1)}{6} \ge \cQ$. As discussed in \cite{baredef}, the scalar mode of the supersymmetric setup is always stable, but the non-supersymmetric setup requires $f_7^2 > 49/32$ for its mass to be above the BF \cite{BF} bound. This stabilization becomes inconsistent, however, in the light of backreaction, as the D7 brane cause an asymptotic deficit angle of $N_f/12$, and we would require $N_f \gg 1$ for a finite value of $f_7$. Hence, we do not follow the path of the massive D3-D7 embeddings that are distinct from the D3-D5 case. Just for curiosity, this instability is reflected in the asymptotic solution for $\Psi$, which becomes in the non-supersymmetric case
$\Psi(u) \sim  u^{\alpha_\pm}$ with $\alpha_\pm = \frac{3}{2}\pm \sqrt{\frac{4Q^2 -7 -12Q}{2+4Q}}$ -- raising interesting questions about the nature of the dual ``mass'' operator. Here, we see that satisfying the BF bound for $\Psi$ precisely corresponds to real values of $\alpha_\pm$, i.e. to a non-oscillatory solution, and the unstable solution would be non-physical. In \cite{quantumhall}, it is, however, given some interpretation in the context of the quantum Hall effect.
\section{Thermodynamics of the defect}\mlabel{thermdyn}
%
Now, we are ready to compute the contribution of the defect to the extrinsic
thermodynamic quantities that are localized on and around the
defect, defined uniquely such that we assume the {\it bulk} contribution to be translationally invariant anywhere else than at the defect, where it is at most discontinuous, and the defect contribution to vanish far away from the defect.
This allows us to study the phase diagram of the matter, the level of stability of the different phases, and their physical properties.
%

Naturally, we will implicitly consider the density
of these quantities per unit area of the defect in terms of the
boundary metric, which removes the divergence from the infinite
volume factor of the integral. We will miss however any contribution to possible changes to
the asymptotic characteristics of the 3+1 SYM. Since the asymptotics however should only depend on the topological properties of the defect, i.e. the flux parameter $f$ and not the local properties of the embedding, this is still sufficient to map out the phase diagram and discuss its properties.

As pointed out in section \mref{fancyintro}, our discussion of the massive embeddings is limited to the D5 case.

\subsection{Free Energy and Thermodynamic Variables}\mlabel{thermdefs}
As a starting point, we can straightforwardly compute the free energy via the standard procedure from the Euclidean action, $I_e$, \cite{long} using 
\begin{equation}\mlabel{freeendef}
F \ = \ T I_e \  ,
\ \ \ I_e \ = \  \int_{u_{min}}^{u_{max}} \mathcal{L}_e \ + \ I_{bdy} \ ,
\end{equation}
where the boundary terms
\begin{equation}
I_{bdy.} \ = \ - \frac{1}{3} \sqrt{\gamma} + \frac{1}{2}\Psi^2
\end{equation}
are dictated to us by consistency \cite{skenderis}. The other boundary terms of \cite{skenderis} do not contribute in our case because of isotropy in the flat directions and rotational symmetry on the sphere. Since the determinant of the boundary metric, $\gamma$, vanishes on the horizon, only the asymptotic boundary contributes. In the case of non-blackhole (Minkowski) embeddings there is only one (i.e. the asymptotic) boundary, provided the action is consistent at $u_{min}$ and the point at $u_{min}$ is included, as will be discussed below.

In the next step, one can construct other thermodynamic quantities, such as the entropy, $S = - \left. \frac{\partial F}{\partial
T}\right|_V$, the energy $E = -F + TS$ or the heat capacity $c_V = \left. \frac{\partial E}{\partial T}\right|_V$. Since the defect is in thermal equilibrium with the bulk, and also the extrinsic curvature of the horizon in the brane geometry is the same as in the bulk theory, the choice for the temperature is obviously the bulk temperature $T = \frac{r_0}{\pi L^2}$. There is a slight ambiguity as to what one considers to be the thermodynamic volume $V$.
One could either consider the defect as an isolated thermodynamic system, embedded in the SYM heat bath, or as part of an overall system. In the former case the volume is either the 2-dimensional volume $\int dx\, dy$ with the effective width of the defect considered to be an ``internal'' degree of freedom or alternatively the 3-dimensional volume $\int dx\, dy \, dz$ over a finite width $\Delta z$, e.g. $\Delta z = \int z'(u) \, du$, with $z'(u)$ given in \reef{backgd}. In the latter case, however, one considers a large volume of 3d SYM plus the defect, and the quantities that we are studying are just the contributions that are extrinsic in the two dimensions of the defect, and independent of the extension of the volume in the $z$-direction -- in the limit of placing the boundary of the volume far away from the defect. To make this more explicit, we can look at the variation of the euclidean action, which gives
\begin{equation}
\delta I_e \  = \frac{\partial \mathcal{L}_e}{\partial \Psi'} \delta \Psi + \frac{\partial \mathcal{L}_e}{\partial A_t'} \delta A_t + \frac{\partial \mathcal{L}_e}{\partial \chi'} \delta \chi \ ,
\end{equation}
or in terms of field theory quantities 
\begin{equation}
\delta F \ = \ 2 C \delta M_q \  +  \ \rho \delta \mu \ + \ F_f \delta z_{max} \ .
\end{equation}
In principle we might expect a term $\propto m \delta m$ that would be divergent. This term however cancels because of the renormalization. Also the variation of the magnetic field does not contribute because of its tensor structure. $F_f$ is just defined as the change of the 3+1 free energy density $F_{SYM}$ due to varying $N_c \rightarrow N_c + q$, in dimensionful units:
\begin{equation}
F_f \ = \ q \frac{\delta F_{SYM}}{\delta N_c}  \ = \ - q \frac{\pi^2}{4} N_c N_f  T^4 \ = \ - \frac{\pi}{4} f N_c N_f \sqrt{\lambda} T^4 .
\end{equation}
If $f \sim \order (1)$, then term is suppressed by a factor $\lambda ^{-1/2}$ compared to $F_{SYM}$. Noting the fact that the SYM background is isotropic such that the pressure equals the free energy, $F_{SYM} = P_{SYM}$, we could interpret $F_f \delta\,  \Delta z$ as a work term for the case of the isolated defect. This demonstrates nicely that this case is inconsistent, since want to consider the defect system ``on-shell'' and study thermodynamic processes obviously with a fixed gauge group, i.e. at constant $f$. This implies that we cannot use $z_{max}$ as an independent thermodynamic variable. Hence, we need to do a change of variables in the thermodynamic potentials, corresponding to a Legendre transformation of the action. Since  we also want to consider processes at fixed baryon density on the defect, rather than at fixed chemical potential, as in \cite{findens}, we do Legendre transformations in $\chi$ and $A_t$:
\begin{equation}
\tilde{\mathcal{L}}_e \ = \ \mathcal{L}_e \ + \ \rho A'_t \ - \ f \chi' \  \ \mathrm{or} \ \ \ \tilde{I}_e \ = \ I_e \ + \ \rho A_t \ - \ f \chi \ .
\end{equation}
Now, the variation of the free energy is 
\begin{equation}\labell{freenvary}
\delta F \ = \ 2 C \delta M_q \  +  \ \mu \delta \rho \ + \ z_{max} \delta F_f \ ,
\end{equation}
which implies $F = F(M_q,\rho, f, T)$. Obviously we consider $f$ fixed, even though it is not inconceivable to have processes in condensed matter physics that change the effective gauge group.

In the case of Minkowski embeddings, the significance of the Legendre transformation can be seen nicely from the brane tension in the $z$ direction at the endpoint of the brane $u_{max}$
that can be straightforwardly computed from
\begin{equation}
\tau_z \, = \, g_{zz}\frac{\delta \mathcal{L}^{(D5)}}{\delta  g_{zz}} \, = \,  g_{zz} z'^2\frac{\delta \mathcal{L}^{(D5)}}{\delta g_{rr}} \ ,
\end{equation}
where $g$ is here the D3 background metric.
This can be shown upon substitution of \reef{backgd} and the boundary condition $\Psi \sim 1 - (u_{max} - u)\frac{1}{u_{max}(1+u_{max}^4)}$ to match precisely the tension of $q  N_f$ D3 branes.

In particular, we can look at the source term that corresponds to attaching an appropriate stack of D3 branes in the flat directions at the endpoint of the probe branes and balances this tension to allow for a consistent static embedding, 
\begin{equation}
\mathcal{L}_{source}\!\!(u) \ =  \  - f\ z(u)\delta(u_{max} - u) \ .
\end{equation}
After integration in the radial direction, this term gives precisely the same contribution to the action as the term $- \ f \chi'$ that we added for the Legendre transform. An interesting comment to add is that in this Minkowski embedding, the radial tension $\tau_r$ vanishes at the endpoint, and is entirely generated by the 5-form flux acting on the brane.
Thus, it is actually possible in this defect setup to construct an at least metastable Minkowski embedding at finite baryon density within string theory -- in contrast to the setups of fundamental matter in (3+1) dimensions in \cite{findens}.
\subsection{Dual potentials and response functions}\mlabel{responses}
As some physical quantities of relevance, we will obtain the entropy $S =  - \left. \frac{\partial F}{\partial T} \right|_V$, the total energy $E = F + T S$, heat capacity $c_V = \left. \frac{\partial E}{\partial T}\right|_V = - \left. \frac{\partial^2 F}{\partial T^2}\right|_V$, baryon number chemical potential $\mu = \left. \frac{\partial F}{\partial \rho} \right|_{T,V}$ and magnetization $M = \left. \frac{\partial F}{\partial B} \right|_{T,V}$. 

The latter two quantities are straightforward, since we only need to keep in mind the temperature scaling and normalization, such that 
\begin{eqnarray}\labell{chemdef}
\mu & = &  \frac{\partial F}{ \partial \rho}\, = \,  \frac{4 \pi }{\sqrt{\lambda}N_c N_f T^2} \frac{\partial F}{ \partial \tlrho} \, =: \, 4\pi T \tilde{\mu} \ \ \mathrm{and} \\ \labell{magdef}
M & = &  \frac{\partial F}{\partial B} \, = \, \frac{1}{(\pi T)^2} \frac{\partial F}{\partial \tlb} \, =: \, \frac{\sqrt{\lambda}N_c N_f T}{ \pi^2 } \tilde{M}\ .
\end{eqnarray}
In principle, there would also be a contribution from the variation of the embedding $\Psi(u)$ via a term $\frac{\delta I_e}{\delta \Psi} \partial_{\rho, B}\Psi$, however, it turns out that it does not contribute for the following reason: Since we did a Legendre transform in $(A'_t, \rho)$ and $(\chi',f)$, the variation $\frac{\delta}{\delta \Psi}$ at constant $f$, $B$ and $\rho$ is on-shell, i.e. only a boundary term contributes:
\begin{equation}
\frac{\delta I_e}{\delta \Psi} \delta \Psi \ = \ \left[\left(\frac{\partial \mathcal{L}_e}{\partial \Psi'}  \ + \ \frac{\partial I_{bdy}}{\partial \Psi}\right)\delta \Psi \right]_{bdy.} \ .
\end{equation}
In principle, this term depends on $\tlm$ and $\tlc$. Keeping $\tlm$ fixed, $\tlc$ will generically depend on $B$ and $\rho$ and hence $\delta \Psi  = \left(\partial_{\tlrho,\tlb} \, \tlc\right) \, u^2 \delta (\tlrho,\tlb)$. However, it turns out that in an expansion around $u=0$ and ignoring overall factors, we have $\frac{\partial \mathcal{L}_e}{\partial \Psi'}  +  \frac{\partial I_{bdy}}{\partial \Psi} \sim \frac{\tlc}{u} + \tlm^2 $, such that any such term will not contribute in the limit $u \rightarrow 0$.
For black hole embeddings, there is obviously also the second boundary at the horizon, but this contribution vanishes since the boundary metric vanishes, $\gamma \rightarrow 0$, as $u \rightarrow 1$. 
In the case of Minkowski embeddings, there exists only one boundary (the asymptotic one), as we learned above that the Legendre-transformed action is consistent at the endpoint $u_{max}$ and corresponds to including this point in the integral. To demonstrate that there is indeed no contribution from the region at $u_{max}$, we can use the expansion \reef{umax} of $\Psi$ near $u_{max}$.
This relates any change $u_{max} \rightarrow u_{max} + \delta u_{max}$ to some $\delta \Psi$. If we then were to exclude the point at $u_{max}$ and evaluate the integral only up to some $u_\epsilon = u_{max} - \epsilon$, the contribution from any on-shell variation of the scalar $\Psi$, $-\left.\frac{\partial \mathcal{L}_e}{\partial \Psi'}\right|_{u=u_\epsilon} \delta \Psi(u_\epsilon)$ cancels the corresponding extra contribution to the integral, $\mathcal{L}_e\!(u_\epsilon) \frac{\delta \Psi(u_\epsilon)}{\Psi'(u_\epsilon)}$ as we take $\epsilon \rightarrow 0$.

Computing the entropy requires a few more steps. In order to compute the temperature derivative, we proceed by computing the integral \reef{freeendef} in terms of the dimensionless versions of the coordinates and fields. We then consider only the indirect temperature dependence of the terms in the integral and an overall explicit temperature factor which we find by dimensional analysis to be $T^3$. Keeping $\rho$, $B$, $M_q$ and $f$ fixed, the temperature dependencies that we will need are given by $\left.\frac{\partial \tlrho}{\partial T}\right|_{\rho}\! = 2 \frac{\tlrho}{T}$, $\left.\frac{\partial \tlb}{\partial T}\right|_{B}\! = 2 \frac{\tlb}{T}$ and $\left.\frac{\partial \tlm}{\partial T}\right|_{M_q}\! = - \frac{\tlm}{T}$.
The variation of the action with respect to $\tlrho$ and $\tlb$ are straightforward and defined in \reef{chemdef} and \reef{magdef}, and the variation with respect to $\Psi$ gives on-shell 
\begin{equation}
\delta \Psi \frac{\delta \, I_e}{\delta \Psi} \ = \ \frac{\partial \mathcal{L}_e}{\partial \Psi'} \delta \Psi \ + \ \int E_\Psi \delta \Psi \ + \ \frac{\partial I_{bdy}}{\partial \Psi} \delta \Psi \ = \ - \frac{2^{2/3}}{\sqrt{\lambda} T} \frac{c}{u_{min}} \delta \Psi|_{u_{min}}
\end{equation}
where $E_\Psi$ is the equation of motion for $\Psi$. Hence, in the limit $u_{min}\rightarrow 0$, only the linear term in $\delta \Psi$ contributes and we do not have to worry about the non-trivial temperature dependence of the condensate.
Finally, we have to worry about the temperature dependence of the boundary of the integral. The boundary at the horizon is fixed in terms of the dimensionless coordinate at $u=1$, so there is no contribution from the horizon. For the asymptotic boundary, we can compute, in dimensionless variables and ignoring overall factors:
\begin{equation}
\left. \frac{\partial I_e}{\partial u_{min}} \right|_{u=u_{min}} \!\! = \  \mathcal{L}_e \ + \ \partial_{u_{min}} I_{bdy.} \ = \ -\frac{1}{6}+\frac{1}{2}\left(\tlrho^2 + \tlb^2 + f^2 + c^2 - \frac{\tlm^4}{4} \right) +\order(u_{min})
\end{equation}
Since we take the limit $u_{min} \rightarrow 0$, any temperature dependence at fixed maximum dimensionful radius $r_{max}$ is proportional to $\partial_T u_{min} = \frac{u_{min}}{T}$ and hence, there is no contribution from this boundary either.
Putting all the non-vanishing contributions together, we arrive with 
\begin{eqnarray}\labell{entget}
 S & = & - \frac{3 F}{T} \, - \,  \frac{2}{T} \left( \frac{\partial F}{\partial \tlrho} \tlrho + \frac{\partial F}{\partial \tlb} \tlb  \right) \, + \, \frac{\delta I_e}{\delta \Psi} \Psi_{bdy.} \ \ \mathrm{or} \\
T\, S & = & - 3 F \, - \,  2 \mu \rho \, - \, 2 M\, B \, + \, M_q \, C \ .
\end{eqnarray}

Finally, we can compute the total energy, $E = F+TS$ and the heat capacity
\begin{equation}
c_V\ =\ \left. \frac{\partial E}{\partial T}\right|_V \ = \ - T \left. \frac{\partial^2 F}{\partial T^2}\right|_V \ .
\end{equation}
The second temperature derivative is computed in the same way as the first derivative for the entropy. The most straightforward way is to take the temperature derivative of the entropy. Keeping in mind that the total temperature derivative of the implicitly present equations of motion $E_\Psi$ vanishes, the remaining terms are:
\begin{eqnarray}
\frac{d^2 \, F}{d\, T^2} \!&\!\! =&\!\! \left(\partial_T + \frac{2 \tlrho}{T}\partial_{\tlrho} +\frac{2 \tlb}{T}\partial_\tlb + \frac{d \, \Psi}{d\, T}\frac{\delta}{\delta \Psi} \right)\left(\partial_T + \frac{2 \rho}{T}\partial_{\rho} +\frac{2 \tlb}{T}\partial_\tlb  \right) F
\, - \,  \frac{d}{d\,T}  \frac{M_q \, C}{T} \\
\!\!\!\!&\!\! = & \!\!\!\left(\partial_T + \frac{2 \tlrho}{T}\partial_{\tlrho} +\frac{2 \tlb}{T}\partial_\tlb  \right)^2\! F  -  3 \frac{  M_q \, C}{T^2}
 + \left( \frac{2 \tlrho}{T}\partial_{\tlrho} +\frac{2 \tlb}{T}\partial_\tlb  \right) \left( \int\!\! E_\Psi \frac{d\Psi}{d \, T}\right)
 - \,  \frac{d}{d\,T}  \frac{M_q \, C}{T} \ , \nonumber
\end{eqnarray}
where the derivatives are considered to act on anything towards their right.
Here, we find that we cannot avoid computing the temperature dependence of the embedding $\Psi$ and the  condensate. 
This can be done in closed form, without resorting to numerical derivatives. To do so, we expand the temperature derivative of the equations of motion for $\Psi$ 
to first order in temperature, obviously keeping the appropriate dimensionful quantities fixed. The term linear in temperature then gives us an inhomogeneous linear second order equation for $\frac{d\, \Psi}{d\, T}$
\begin{equation}
\tlb\frac{\delta \, E_\Psi}{\delta \, \tlb}\, +\, \tlrho\frac{\delta \, E_\Psi}{\delta \, \tlrho}  \ = \ \frac{T}{2}\left(  \frac{\delta \, E_\Psi}{\delta \, \Psi}\frac{d\, \Psi}{d\, T}\, +\,\frac{\delta \, E_\Psi}{\delta \, \Psi'}\frac{d\, \Psi'}{d\, T}\, +\,\frac{\delta \, E_\Psi}{\delta \, \Psi''}\frac{d\, \Psi''}{d\, T}\right)
\end{equation}
and a boundary on the horizon condition fixing $\left(\frac{d\, \Psi}{d\, T}\right)^{-1} \partial_u \left(\frac{d\, \Psi}{d\, T}\right)$ - both of which are not very illuminating and we do not explicitly write them out here. 
Finally, we choose the boundary condition
\begin{equation}
\partial_u \left[\frac{d\, \Psi}{d\, T}\right]_{u_{min}} \ = \ - \left. \Psi'\right|_{u_{min}}
\end{equation}
at the asymptotic boundary to account for the temperature derivative of the dimensionless mass parameter -- giving the problem numerically slightly non-trivial mixed boundary conditions.
Alternatively, the heat capacity can also be written in a more systematic way,
\begin{eqnarray}
c_v & = & - T \left(\partial_T \, +\,  2 \frac{B}{T} \frac{d}{d\, B}\,+\,  2 \frac{\rho}{T} \frac{d}{d\, \rho} \, -\,   \frac{M_q}{T} \frac{d}{d\, M_q} \right)^2 F \\ \nonumber
& = & - 4 B^2\frac{\chi_B}{T}\,  - \,  4 \frac{\rho^2}{\epsilon\, T} \, - \, 8 \rho B \frac{m}{T} \, + \, \frac{M}{T} \left(4 B \frac{d}{d\, B}\,+\,  4\rho \frac{d}{d\, \rho} \, -\,   M_q \frac{d}{d\, M_q} \right) C \, - \, 6 \frac{F}{T} \ ,
\end{eqnarray}
where we identified the magnetic susceptibility, density of states, and mean magnetic moment:
\begin{eqnarray}
\chi_B \ = & \left. \frac{d\, M}{d\, B}\right|_{T,V,\rho} & = \ \frac{1}{(\pi T)^4} \int \left(\partial_\tlb^2 \mathcal{L}_e + \partial_\tlb E_\Psi \frac{d\Psi}{d \tlb} \right) \\
\epsilon^{-1} \ = & \left.\frac{d\, \mu}{d\, \rho}\right|_{T,V,B} & = \ \frac{16 \pi^2 }{ \lambda N_c^2 N_f^2 T^4}  \int \left(\partial_{\tlrho}^2 \mathcal{L}_e + \partial_{\tlrho} E_\Psi \frac{d\Psi}{d \tlrho} \right) \\
m \ = & \left. \frac{d\, M}{d\, \rho}\right|_{T,V,B} & = \ \frac{4  }{\sqrt{\lambda}N_c N_f T^4} \int \left(\partial_\tlb \partial_{\tlrho} \mathcal{L}_e + \partial_{\tlrho} E_\Psi \frac{d\Psi}{d \tlb} \right) \ .
\end{eqnarray}
The derivatives of the scalar $ \frac{d\Psi}{d \tlb}$ and $\frac{d\Psi}{d \tlrho}$ can be computed in an equivalent fashion as the derivative $\frac{d \Psi}{d\, T}$ described above.


\subsection{Massless case}\mlabel{massless}
In the massless case, the free energy can be integrated straightforwardly analytically and becomes 
\begin{eqnarray}
F & = & - 4\pi T_5 r_0^3 \int_0^1 du \left(\frac{u^4 -3}{3u^4 \sqrt{1-u^4}} + \frac{\sqrt{1 + \left(f^2 + \tlrho^2 + \tlb^2\right)u^4}}{u^4} \right) \nonumber \\
& = & - \sqrt{\lambda}  T^3
N_c N_f \frac{1}{3}\left(\sqrt{1 + f^2 + \tlrho^2 + \tlb^2} \right. \  \\ \nonumber
& & ~~~~~  \left.+ \ 2 \left(-\left(f^2 + \tlrho^2 + \tlb^2\right)\right)^{3/4}\, \mathcal{F}\left(\sinh^{-1}\left(-\left(f^2 + \tlrho^2 + \tlb^2\right)\right)^{1/4}\Big|-1\right) \right)\ ,
\end{eqnarray}
where the first term in the integral cancels the divergence of the second term at $u \to 0$; we substituted field theory quantities in the dimensionful factor in the second line and
$\mathcal{F}(\cdot|\cdot)$ is the incomplete elliptic integral of the first kind. For convenience, we give the asymptotic expansion $\mathcal{F}\left(\sinh^{-1}\left(-X\right)^{1/4}\big|-1\right) = i \mathcal{K}(2) - (-X)^{-1/4} + \order(X)^{-5/4}$ and the expansion at 
\MFIGURE{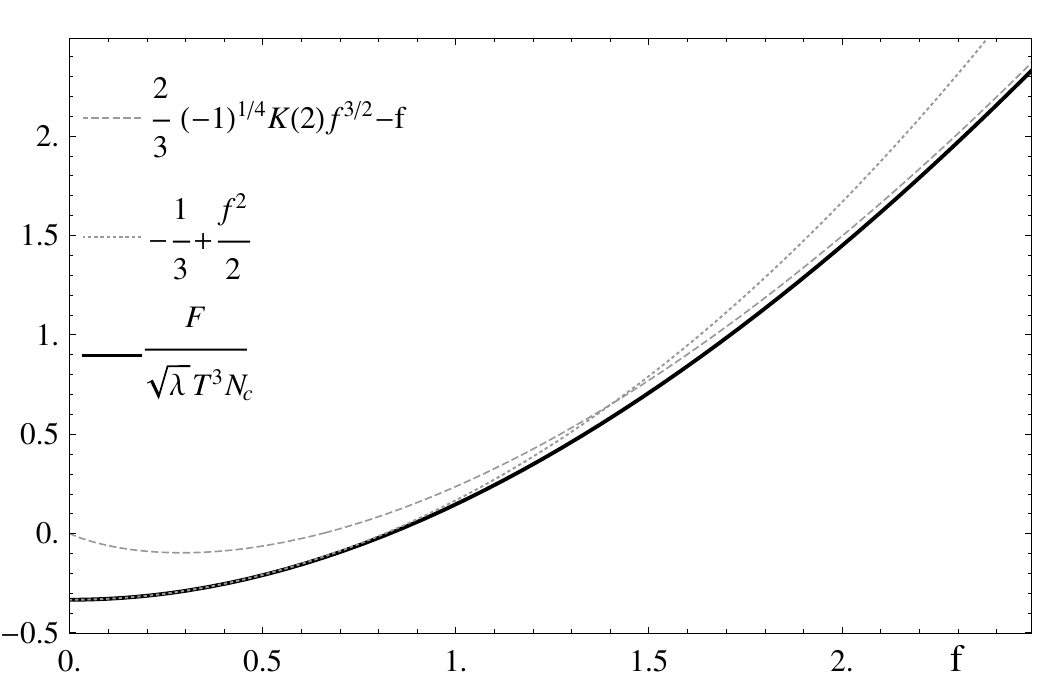}{The free energy density of the defect as a function of $\sqrt{f^2 + \tlb^2+\tlrho^2}$ (denoted for simplicity as ``$f$''.}{freeen_plot_M0}
small values $\mathcal{F}\left(\sinh^{-1}\left(-X\right)^{1/4}\big|-1\right)  = (-X)^{1/4}\left(1-\frac{X}{10}\right) + \order(-X)^{9/4}$, where $\mathcal{K}(.)$ is the complete elliptic integral of the first kind and $ \mathcal{K}(2) \sim 1.854 e^{i \pi/4}$. These asymptotic approximations are indicated in fig. \mref{freeen_plot_M0}. We also note again the effect of the electric-magnetic duality, which relates quantities under the interchange of density and magnetic field.
Now this emerges in the form that the free energy depends only on the variable $f^2 + \tlrho^2 + \tlb^2$. The representation of this symmetry in the various response functions then follows straightforwardly. For example the chemical potential is then related to the magnetization as $\tilde{\mu}(\tlrho, \tlb,f) = \tilde{M}(\tlb,\tlrho, f)$, and in the following we will show only either of them.

Now, we can simply verify the relations \reef{chemdef} and \reef{magdef}:
\begin{eqnarray}\labell{mag_formula}
\tilde{\mu} & = &  \tlrho \frac{\mathcal{F}\left(\sinh^{-1}\left(-\left(f^2 + \tlrho^2 + \tlb^2\right)\right)^{1/4}\Big|-1\right)}{\left(-\left(f^2 + \tlrho^2 + \tlb^2\right)\right)^{1/4}}  =  \sqrt{\lambda}  T^3 N_c N_f  \int_0^1 \!\! du A_t'  \\
z_{max} & = & \left. \frac{\partial F}{\partial F_f} \right|_T \ = \ \frac{4}{\pi N_c N_f \sqrt{\lambda} T^4} \frac{\partial F}{\partial f} \\ \nonumber
 & = & \ \frac{1}{\pi T} f \frac{\mathcal{F}\left(\sinh^{-1}\left(-\left(f^2 + \tlrho^2 + \tlb^2\right)\right)^{1/4}\Big|-1\right)}{\left(-\left(f^2 + \tlrho^2 + \tlb^2\right)\right)^{1/4}} \ = \ \pi T \int_0^1 \! d\, u \, \chi' \ .
\end{eqnarray}
\DOUBLEFIGURE{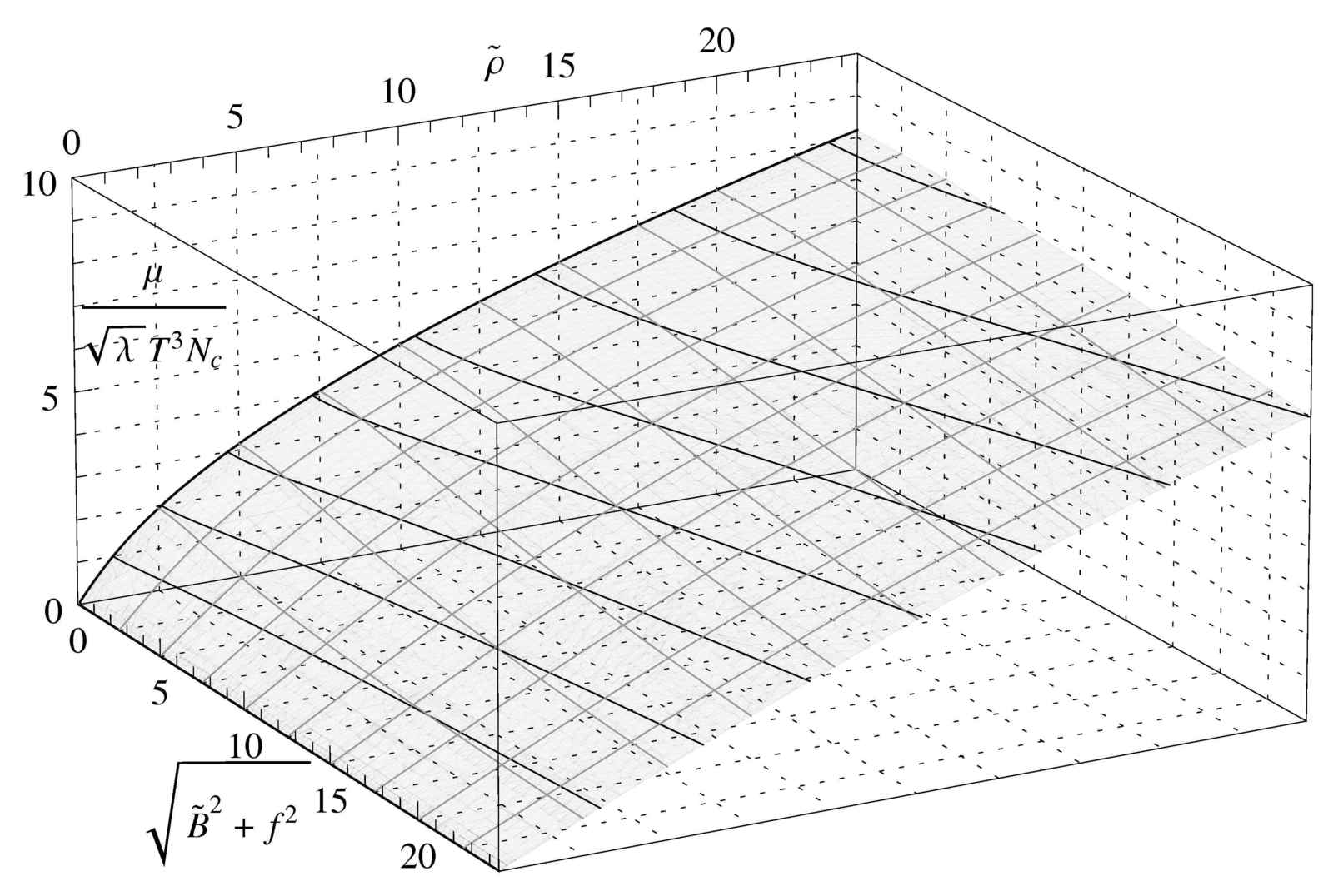,width=0.49\textwidth}{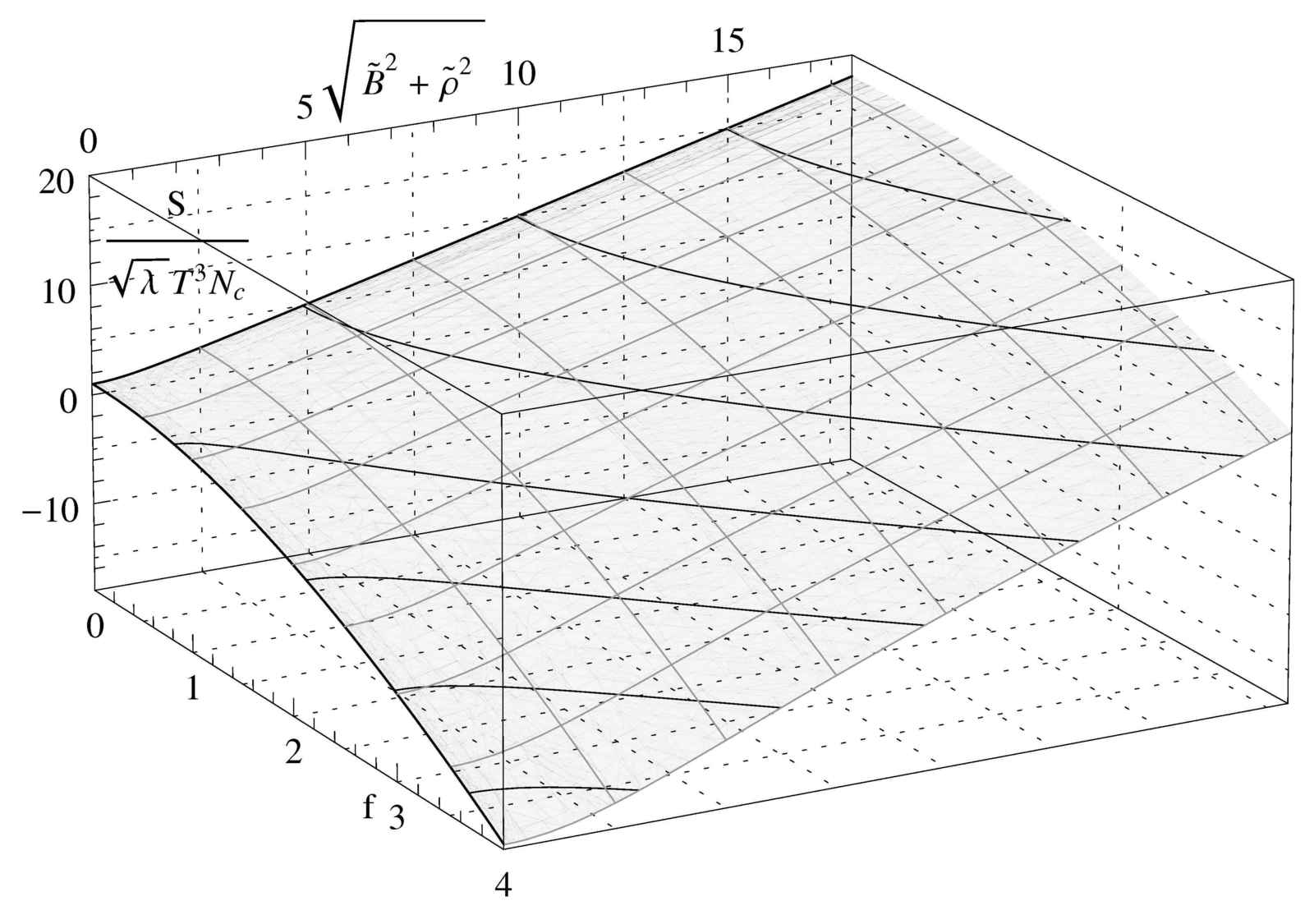,width=0.49\textwidth}{The baryon number chemical potential as a function of $\tlrho$ and $\sqrt{\tlb^2 + f^2}$. Note that, due to electromagnetic duality, this is the same as the magnetization as a function of $\tlb$ and $\sqrt{\tlrho^2 + f^2}$, upon an appropriate scaling with a dimensionful constant.\label{chempot_plot_M0}}{The contribution of the defect to the entropy as a function of $\sqrt{\tlrho^2 + \tlb^2}$ and $f$.\label{entropy_plot_M0}}
This is shown in fig. \ref{chempot_plot_M0}, and we note that the dominant dependence of $\mu$ on $\rho$ or $M$ on $B$, with an apparent saturation behavior as the scaling power changes from $1$ to $1/2$. The dependence on the subleading quantities is a small suppression, which goes slightly against classical intuition.
Further more, we can straightforwardly give the contribution of the defect to the entropy, which is
\begin{equation}
 S  =  \sqrt{\lambda}  T^2 N_c N_f \left(\sqrt{1\! + \! f^2\! +\! \tlrho^2\! +\! \tlb^2}\, +\,  2 f^2 \frac{ \mathcal{F}\left(\sinh^{-1}\left(-\! \left(f^2\! +\! \tlrho^2\! +\! \tlb^2\right)\right)^{\! 1/4}\Big|-1\right)}{\left(-\left(f^2 + \tlrho^2 + \tlb^2\right)\right)^{1/4} } \right) 
\end{equation}
and is shown in fig. \mref{entropy_plot_M0}.
%
%
%
%
%
%
Now, we can notice that for large values of $f$, this expression turns negative. Since the thermodynamic quantities that we derive here are only contributions to the overall quantities of the system ``defect+background'', and the defect is certainly strongly coupled to the background and hence always in thermal equilibrium, this is not troublesome, as we can demonstrate:\\
First, we assume that the defect effectively extends roughly homogeneously over a region up to $z_{max}$ in the normal direction. This is suggested by \reef{backgd}, \reef{mag_formula} and at large $f$ also by the quasiparticle spectrum found in \cite{baredef,fancycon}. Then, we consider the entropy density over this region of $(3+1)$-volume,
\begin{equation}
 \frac{S_{\Delta z}}{z_{max}}  =  2 \sqrt{\lambda} \pi  T^3 N_c N_f \left(
\frac{\left(-\left(f^2 + \tlrho^2 + \tlb^2\right)\right)^{1/4} \sqrt{1 + f^2 + \tlrho^2 + \tlb^2}}{f \mathcal{F}\left(\sinh^{-1}\left(-\left(f^2 + \tlrho^2 + \tlb^2\right)\right)^{1/4}\Big|-1\right)}  -  2 f\right) \ ,
\end{equation}
where we look in particular at the density of the negative term, $ - 2 \pi \sqrt{\lambda}  T^3 N_c N_f f$. In our limit of large $N_c$ and  $N_c \gg f^{2}$, this term is precisely (minus) the contribution $q \frac{\delta S_{SYM}}{\delta N_c}$. This can be interpreted simply in the way that the extra degrees of freedom due to changing $N_c \rightarrow N_c + \delta N_c$ become fully available only after the brane falls into the horizon, and the positive term in the entropy of the defect describes the extra degrees of freedom contributed by the defect inside that region. Furthermore, there is no region in which the total entropy of the combined system is negative. 

At large densities or magnetic fields, or low temperatures, the contribution to the entropy is just
$4 \pi \rho$
%
or 
$\frac{\sqrt{\lambda}N_c N_f B}{4 \pi^2}$
respectively - indicating that the number of degrees of freedom is independent of the temperature, and proportional to the number of quarks or magnetic states.

We can also straightforwardly compute the heat capacity:
\begin{eqnarray}
\!\!\!\!\!\!\!\!\! c_V \ = \ 2 \sqrt{\lambda}  T^2 N_c N_f\left( \frac{1}{\sqrt{1 + f^2 + \tlrho^2 + \tlb^2}}\left(1 + 2 f^2 - \frac{f^4}{f^2 + \tlrho^2 + \tlb^2} \right) \right. ~~~ \nonumber\\
\!\!\!\!\!\!\!\!\! \left. - \, \frac{ f^2 \mathcal{F}\left(\sinh^{-1}\!\!\! \left(-\left(f^2 + \tlrho^2 + \tlb^2\right)\right)^{1/4}\Big|-1\right)}{\left(-\left(f^2 + \tlrho^2 + \tlb^2 \right) \right)^{1/4}}\left(3- \frac{f^2}{f^2 + \tlrho^2 + \tlb^2}  \right)\right) \ ,
\end{eqnarray}
which we show in figure \mref{heatcap_plot_M0}.
%
%
%
\DOUBLEFIGURE{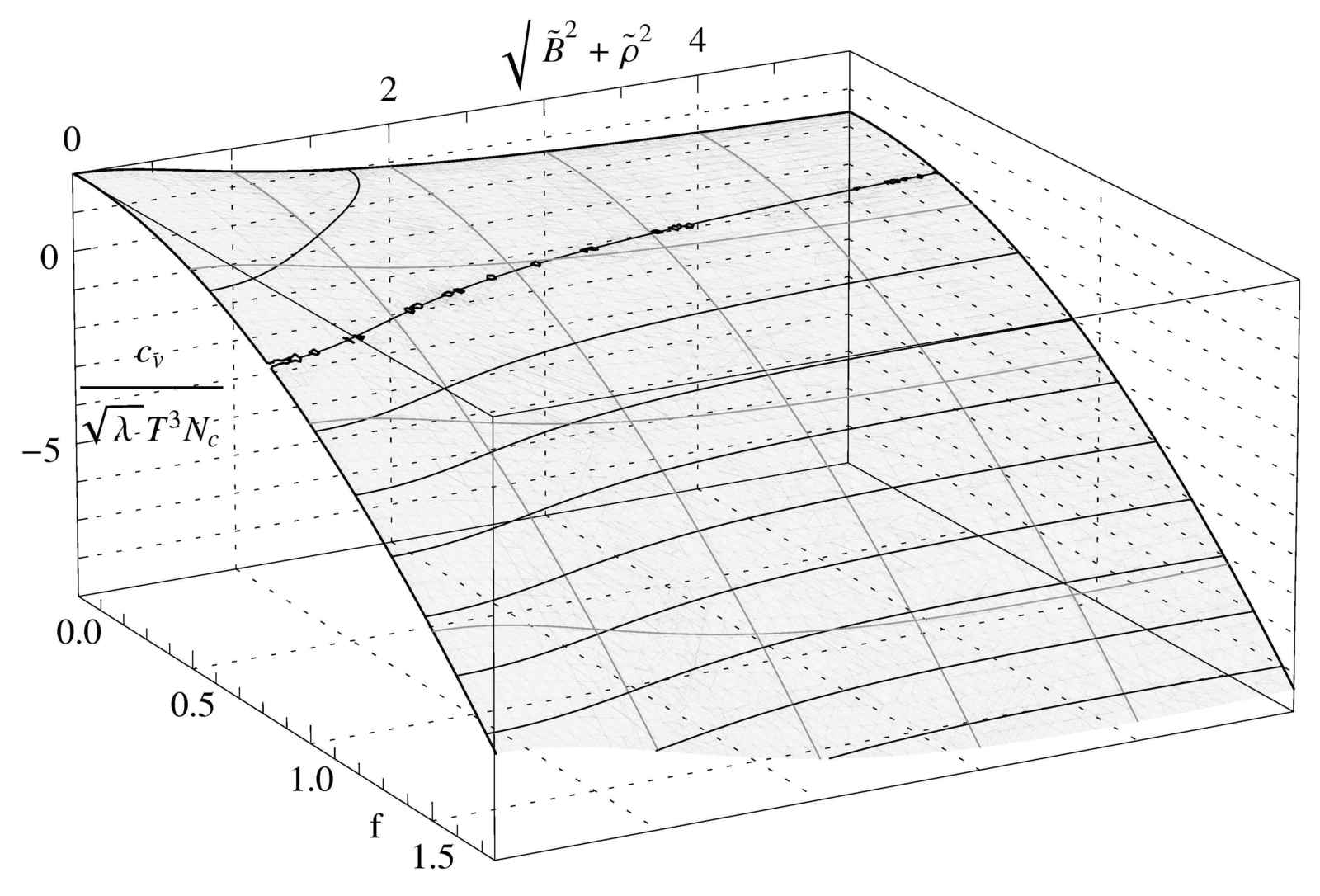,width=0.49\textwidth}{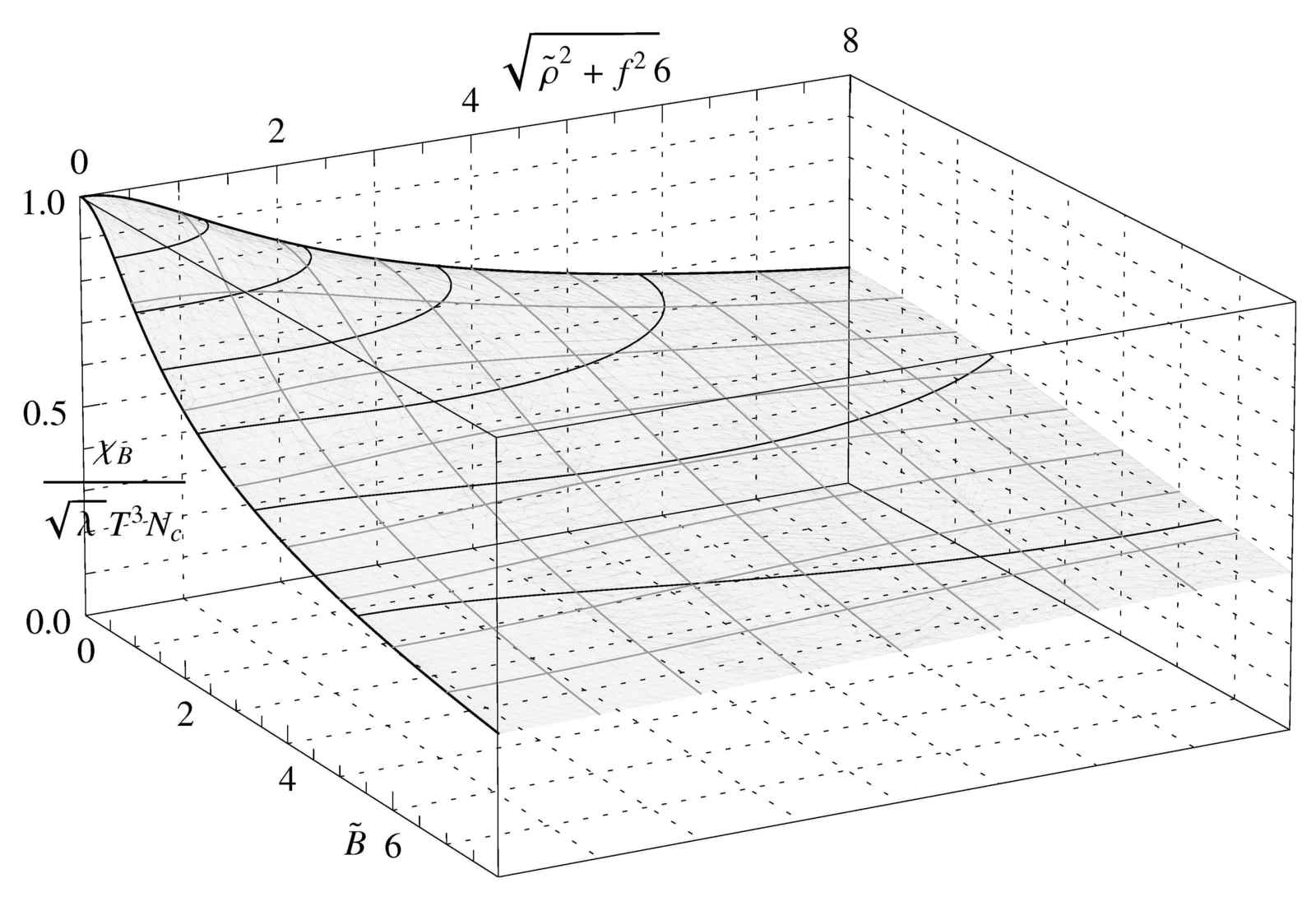,width=0.49\textwidth}{The contribution of the defect to the specific heat as a function of $\sqrt{\tlrho^2 + \tlb^2}$ and $f$.\mlabel{heatcap_plot_M0}}{The magnetic susceptibility of the defect as a function of $\tlb$ and $\sqrt{\tlrho^2 + f^2}$. This is the same as the density of states as a function of $\tlrho$ and $\sqrt{\tlb^2 + f^2}$.\mlabel{suscept_plot_M0}}
Again, we notice that the heat capacity is negative for large $f$, but as in the case with the entropy, this does not signify an instability, as the contribution is much smaller than the heat capacity of the background SYM in the appropriate region around the defect. Also here, the negative contribution to the heat capacity simply indicates that the additional degrees of freedom only turn on gradually over some region $\sim z_{max}$ away from the defect. 
At small $f$, the heat capacity is just $c_V \ = \ 2 \sqrt{\lambda}  T N_c N_f \frac{1}{\sqrt{1 +  \tlrho^2 + \tlb^2}} + \order(f^2)$. This is somewhat counter-intuitive, as one normally expects an increasing heat capacity with increasing density but as the system is strongly coupled, we can only speculate about explanations for this behavior. We have to keep in mind that $\tlrho$ is only the net density and there is always a finite density of quarks and gluons. 

Finally, we can, for example, look at magnetic the susceptibility,
\begin{eqnarray}
\!\!\!\!\!\! \chi_B & = &  \frac{\sqrt{\lambda}  N_c N_f }{2 \pi^2 T } \left(\frac{\tlb^2}{\left(f^2 + \tlrho^2 + \tlb^2 \right)\sqrt{1 + f^2 + \tlrho^2 + \tlb^2}} \right. \\ \nonumber
& + & \left. \left(2 - \frac{\tlb^2}{\tlrho^2+\tlb^2 + f^2 } \right) \frac{  \mathcal{F}\left(\sinh^{-1}\left(-\left(f^2 + \tlrho^2 + \tlb^2\right)\right)^{1/4}\Big|-1\right)}{\left(-\left(f^2 + \tlrho^2 + \tlb^2 \right) \right)^{1/4}} \right) \ .
\end{eqnarray}
This expression is always less than $\chi_B^{(0)} = \sqrt{\lambda}  N_c N_f \pi^2 T^3$ -- the susceptibility of the defect without any of the parameters turned on as shown in fig. \mref{suscept_plot_M0} - and behaves asymptotically as $\chi_B^{(0)} \frac{(-1)^{1/4}\mathcal{K}(2)}{\sqrt{f}}$, $\chi_B^{(0)} \frac{(-1)^{1/4}\mathcal{K}(2)}{\sqrt{\tlrho}}$ or $\chi_B^{(0)} \frac{(-1)^{1/4}\mathcal{K}(2)}{2 \sqrt{\tlb}}$, respectively. Hence, the defect is diamagnetic. Again, we can blame this on the strong coupling, which may increase the spin-spin interactions. Similarly, strong coupling may have the effect of suppressing the energy density of the plasma as we increase the quark density.
%
%
%
%
%
%
\subsection{Massive case}\mlabel{massive}
\subsubsection{Phases}\mlabel{phases}
Now, let us look at the phase diagram in fig. \mref{phases_B_plot}. 
Looking at the different kinds of embeddings, we find that in the case of vanishing $\rho$ and $f$, there are three phases: the stable phases of the blackhole embedding at small masses $M_q < M^{(BH)}_{ max}$ (or large temperatures), denoted as ``{\bf{}B}'' and the Minkowski embedding at large masses $M_q > M^{(flat)}_{ min}$ ({\bf M}). In principle, there exists also an unstable phase with $M^{(BH)}_{ max} > M_q > M^{(flat)}_{ min}$ ({\bf C}), but in practice, this phase is not realized and there will just be a first order phase transition from {\bf B} to {\bf M}.
At finite $f$ or $\rho$, the mass diverges at $u_{max} = 0$ or $\Psi_0 = 1$, so we find two additional phases, each one for the blackhole ({\bf B1}) and Minkowski ({\bf M1}) embeddings. We interpret this {\bf B1} embedding as a continuous deformation of the Minkowski phase with free quarks, but also a high density of mesons. In many situations, the phase transition between those phases will disappear, as we will see below.
Using the condensate as an order parameter to identify different phases, we show the various phases in figs. \ref{phases_B_plot} - \ref{phases_F_plot}. In order to demonstrate the phase transitions, and the disappearance of the blackhole phase {\bf M} and the transition phase {\bf C} in the presence of strong magnetic fields, we plot the results at fixed temperature with varying $M$, rather than fixed mass and varying $T$. Obviously, in a thermodynamic process $M/(\pi T)=0$ cannot be attained.
%
%
\DFIGURE{\begin{tabular}{l r}\includegraphics[width=0.5\textwidth]{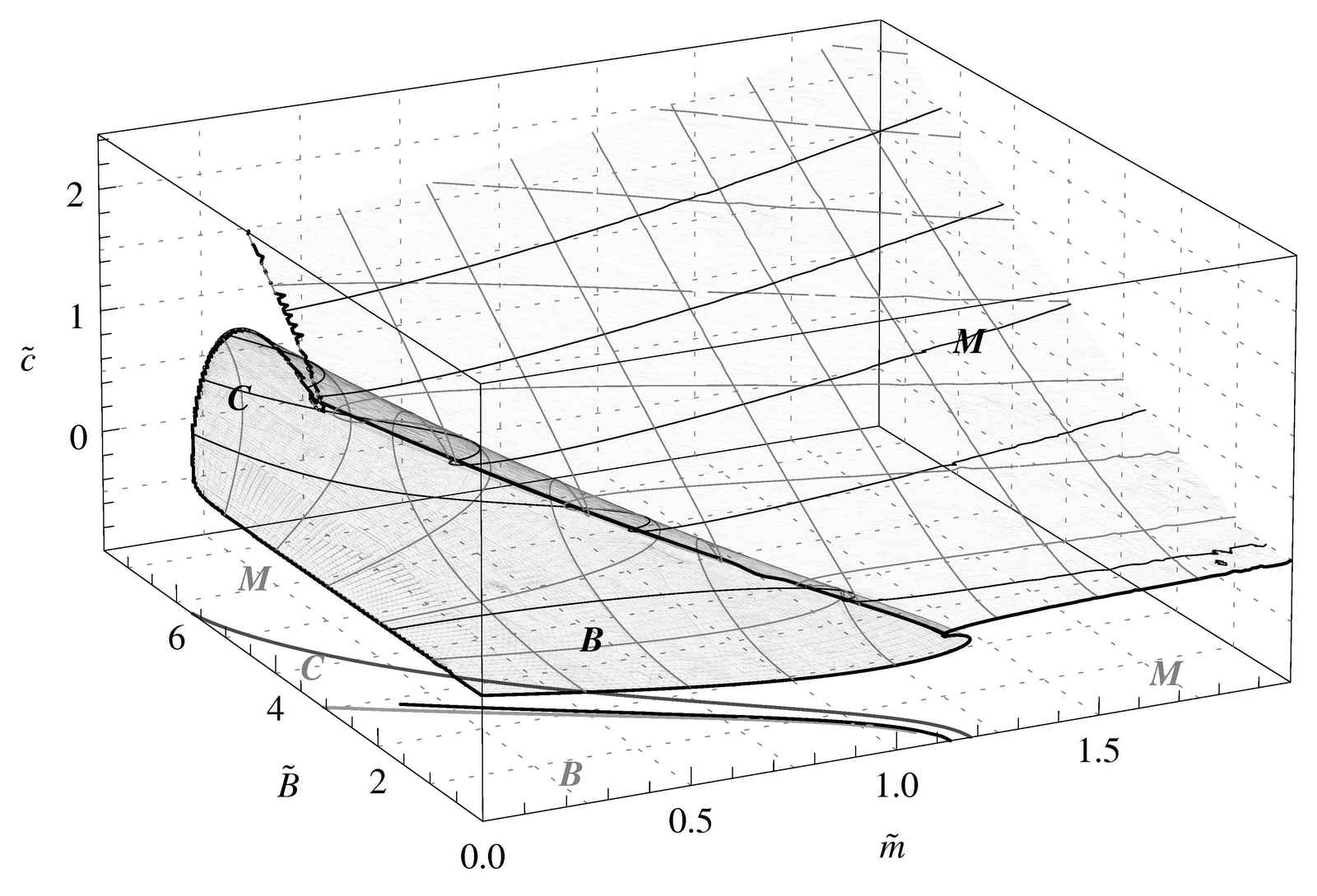}&\includegraphics[width=0.5\textwidth]{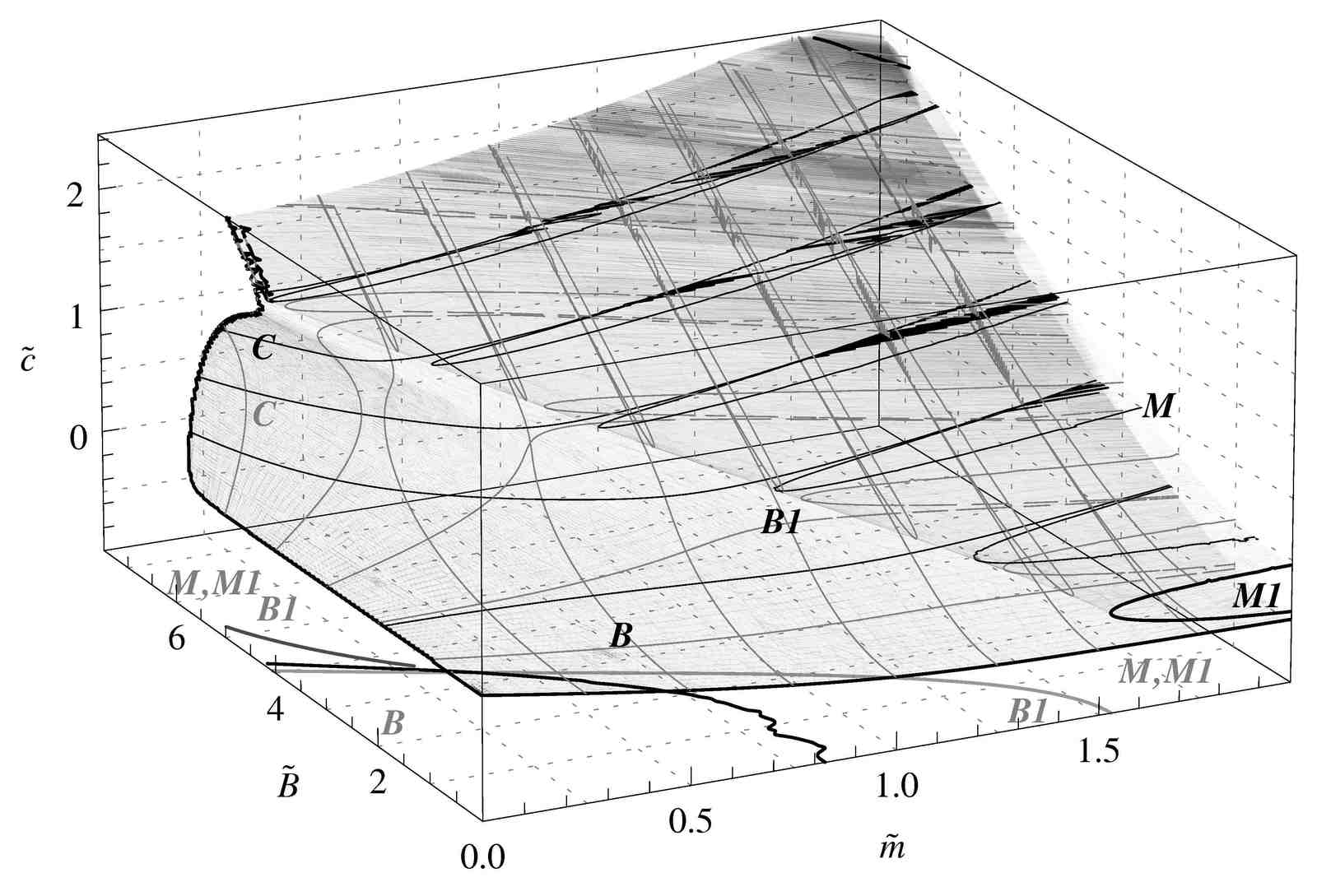}\\
\begin{minipage}{0.49\textwidth}\includegraphics[width=\textwidth]{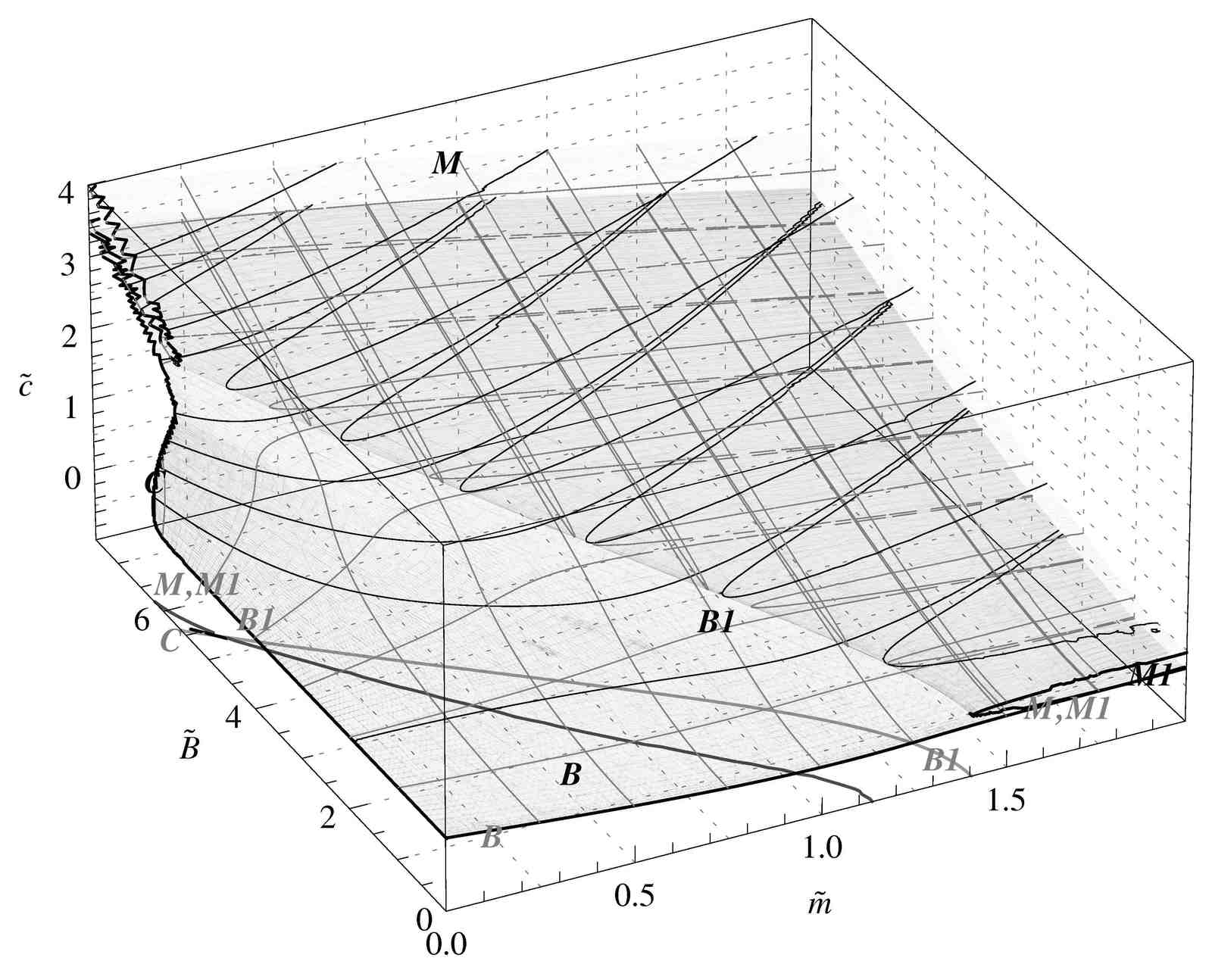}\end{minipage}&
\begin{minipage}{0.49\textwidth}
{\small The light gray (upper) surface is the mesonic, i.e. Minkowski, phase and the darker (lower) surface the blackhole phase. The gray line is the projection of the minimum mass of the Minkowski phase $M^{(flat)}_{ min}$. The black line is the minimum mass of the blackhole phase and the dark gray line the local maximum $M^{(BH)}_{ max}$ , if it exists, or the slowest rate of change of the mass with respect to the embedding. There is some numerical noise visible in the plots that can be ignored, but we chose not to suppress artificially.}
\end{minipage}
\end{tabular}
\caption{Value of the condensate as a function of the dimensionless mass, $\tilde{m}$ and magnetic field, $\tilde{B}$ at fixed temperature.  Top left: Vanishing density and vanishing compact flux $f$. The light and dark gray lines identify the transition between {\bf B} and {\bf C} and {\bf M}, respectively. Top right: Density $\tilde{\rho} = 0.5$. The dark gray line is now the transition between {\bf B} and {\bf C}, or {\bf B} and {\bf B1} and the black line between {\bf C} and {\bf B1}. The light gray line is the transition between {\bf M} and {\bf M1}. Bottom left: $f=0.25$. More details are explained in the bottom right.}\mlabel{phases_B_plot}}
In fig. \ref{phases_B_plot}, we find that the critical mass decreases approximately linearly with the increasing magnetic field and there is a critical magnetic field, above which the blackhole phase {\bf B} disappears. This indicates that the magnetic field catalyzes meson formation. This behavior is similar to what was observed in 3+1 dimensional systems in \cite{magnetic}.
Rigorously speaking, however, there exists at all magnetic fields a continuation of the {\bf B} phase at (exactly) zero mass and we will see in the next section that we can always attain both the Minkowski and blackhole phases at fixed magnetic field and mass and varying temperature.
Also, we find how the phase {\bf C}, that is suppressed and even disappears at finite density or $f$ re-appears at large magnetic fields.
The first order phase transition from {\bf B} to {\bf B1} turns into a smooth crossover as {\bf C} disappears, and one might wonder what happens at the particular point, where {\bf C} disappears. 

The condensate is large in the blackhole phase {\bf B1} and in the Minkowski phases and grows approximately proportional to the magnetic field  -- indicating chiral symmetry breaking. As indicated above, these phases have a very similar behavior. The reader is reminded however that at finite density and vanishing $f$, the Minkowski embedding is not physical within string theory.  It is interesting though that turning on $f$ reverses the linear mass dependence of the  condensate when interpreted as the conjugate potential of the mass -- meaning that if we increase the mass or lower the temperature, the free energy will ``saturate''.
%
In the blackhole phase {\bf B} in which the mesons are dissociated and we have only free quarks,
the condensate approximately vanishes and becomes independent of mass and magnetic field. 

%
%
\DFIGURE{\includegraphics[width=0.5\textwidth]{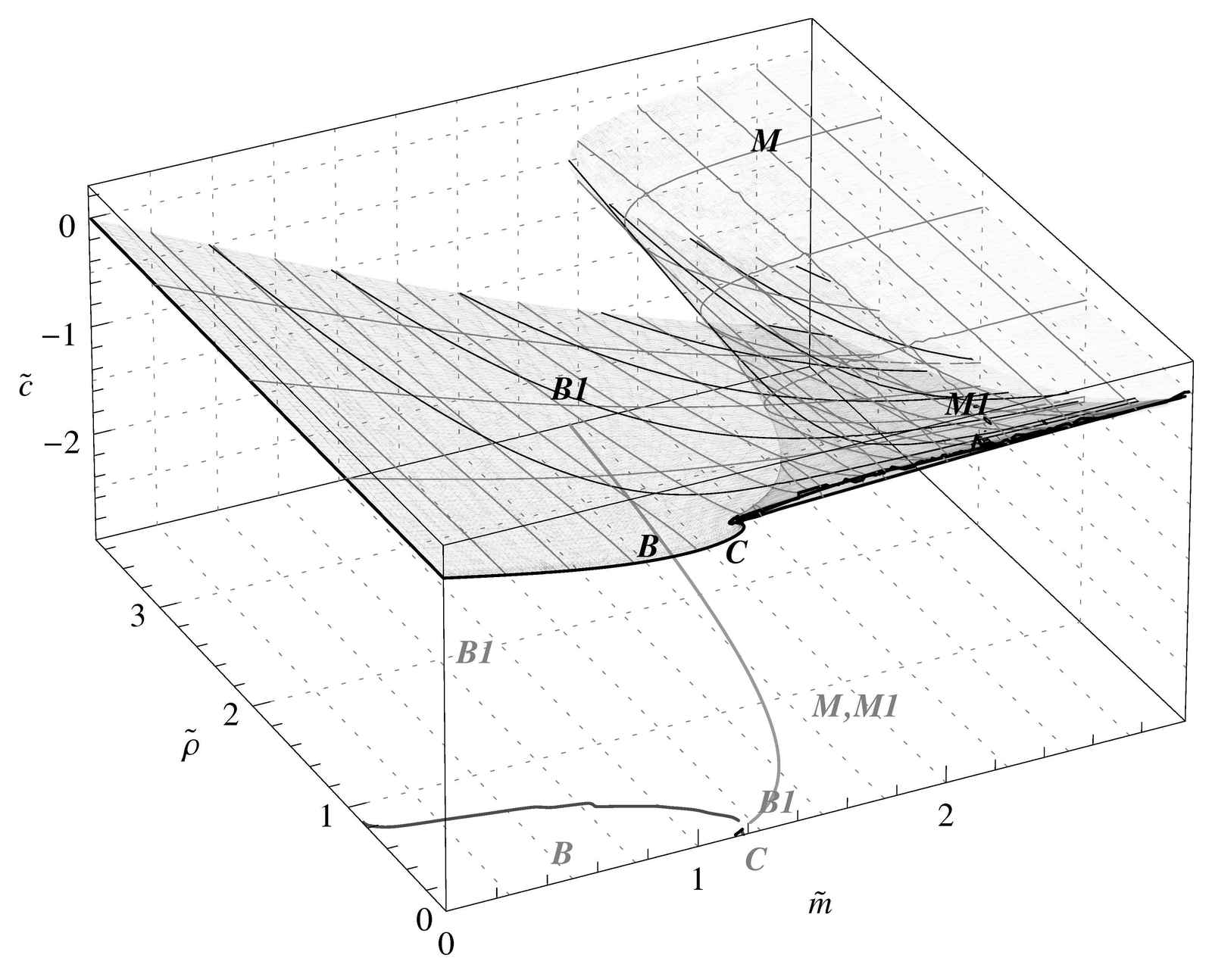}\includegraphics[width=0.5\textwidth]{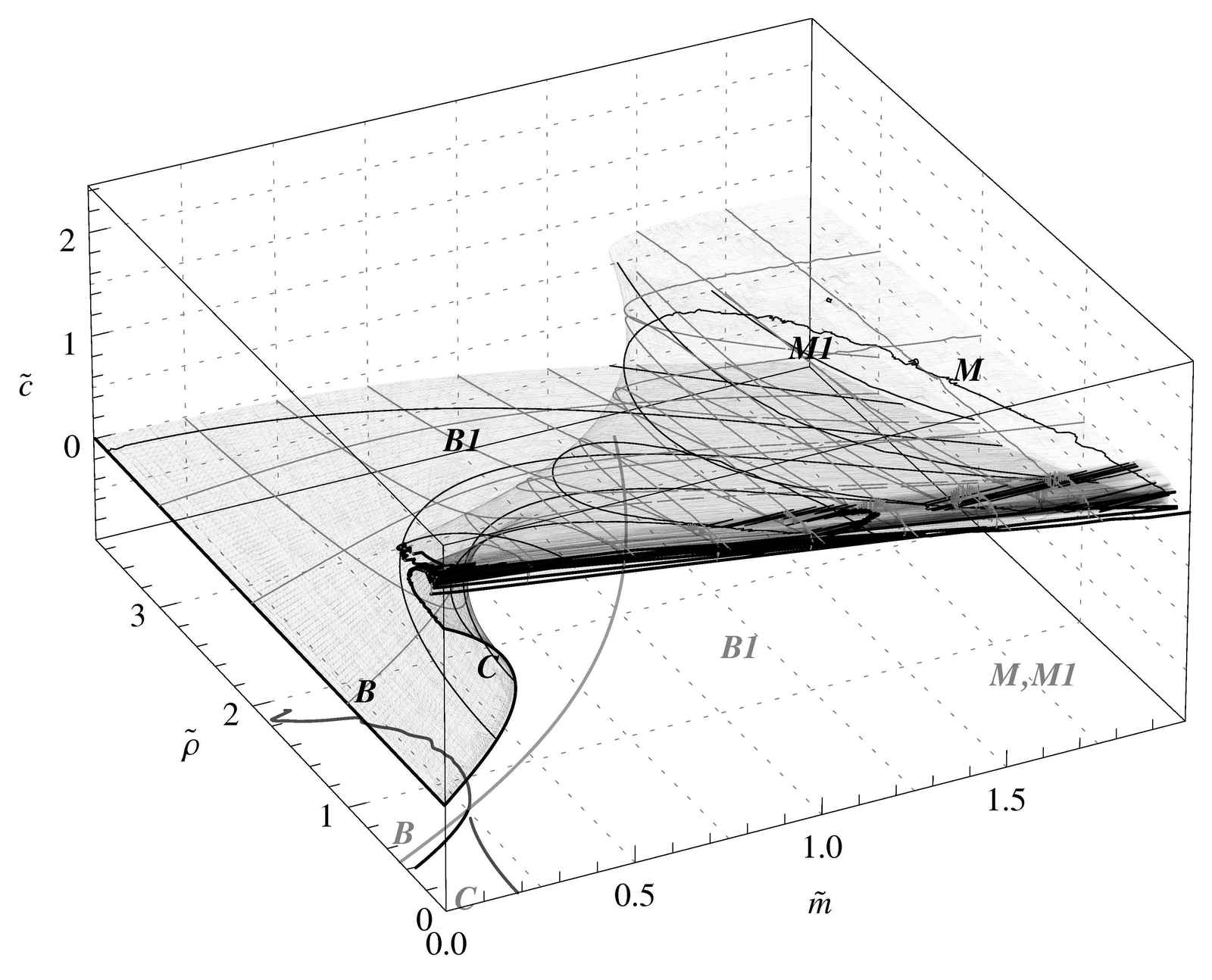}\\
\caption{Value of the condensate as a function of the dimensionless mass and density, at fixed temperature. Left: $\tilde{B}=0$, right $\tilde{B} = 4$.}\mlabel{phases_E_plot}}

In figures \ref{phases_E_plot} and \ref{phases_F_plot}, we see first of all that the phase diagram at finite density and vanishing magnetic field is identical to the one at finite $f$ under identifying $\tlrho \leftrightarrow f$. This can be seen explicitly also from the equation of motion \reef{psieom} of the scalar, which is identical under the exchange $\tlrho \leftrightarrow f$ at $\tlb = 0$. This is however only accidental and disappears as soon as we look e.g. at the value of the action or the free energy and the response functions.
%
%
%
%
\DFIGURE{\includegraphics[width=0.5\textwidth]{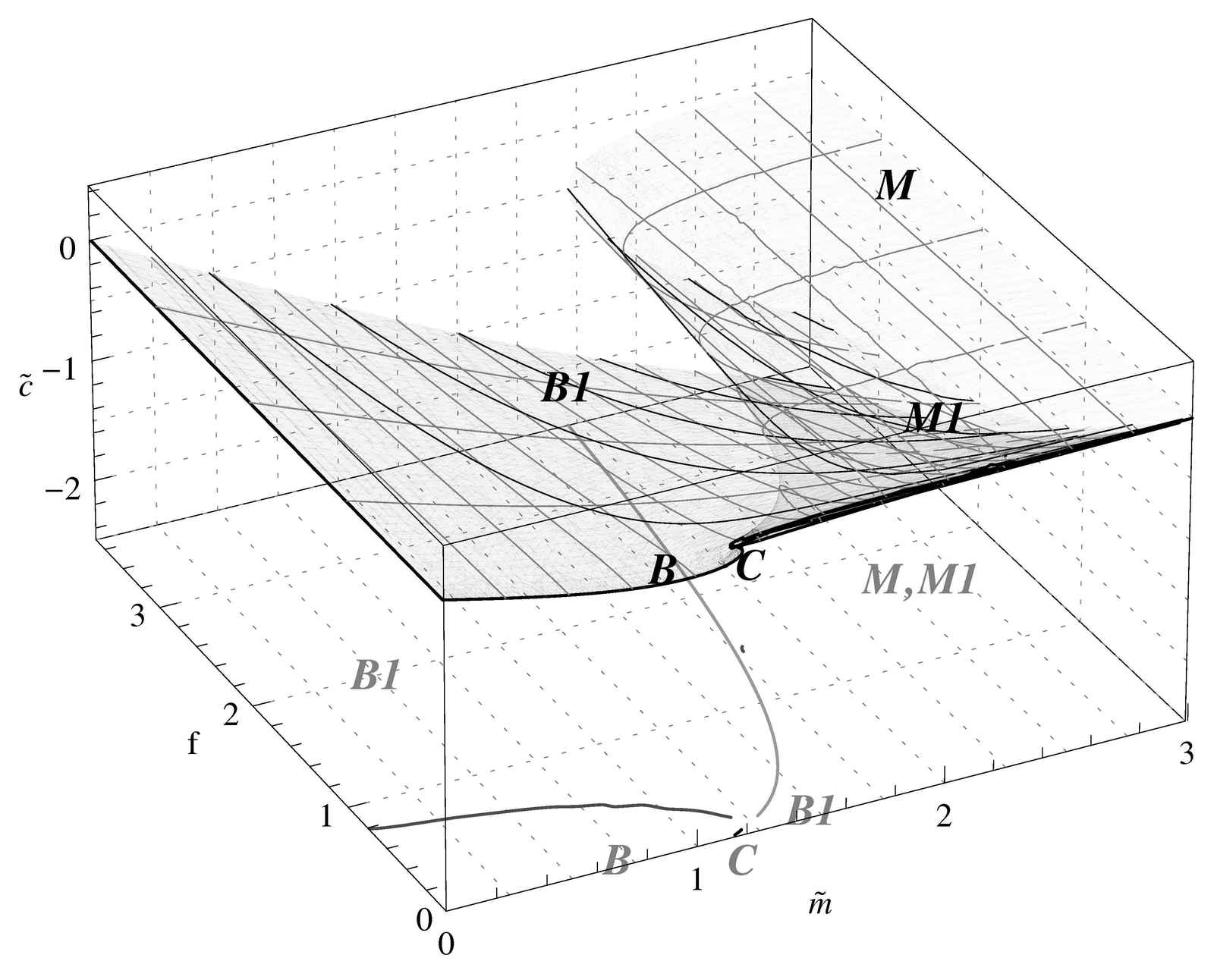}\includegraphics[width=0.5\textwidth]{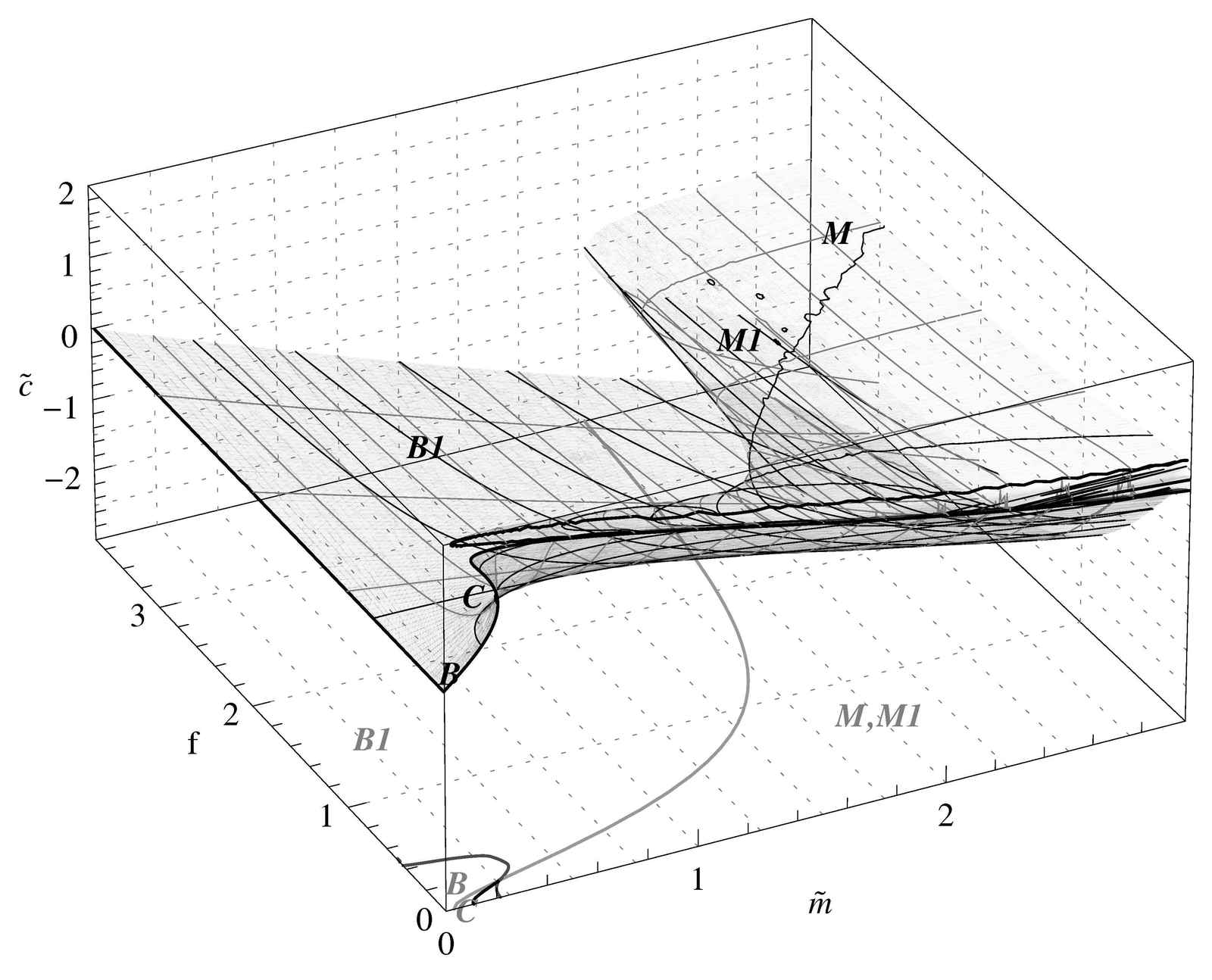}\\
\caption{Condensate as a function of the dimensionless mass and $f$, at fixed temperature.  Left: $\tilde{B}=0$, right $\tilde{B} = 4$.}\mlabel{phases_F_plot}}
Turning on a magnetic field breaks this symmetry even though there is some overall similarity. It has the expected effect of lowering the critical masses,  increasing phase space volume of the phases {\bf C} and {\bf B} and shifting the overall  condensate - breaking also the symmetry of the {\bf M} phase. Interestingly, this persists even at relatively large densities. To understand this behavior we recall that there is always a finite total baryon density, and the polarizing an localizing effect of the magnetic field is not affected by changing the ratio of quarks and anti-quarks.
%
%

\subsubsection{Stability and chemical potentials}\mlabel{pots}
In order to determine which phases are thermodynamically preferred, we can look at the entropy. Before studying the data, we should notice that there is a trade-off between numerical accuracy and noisiness. This arises because a significant contribution to the entropy comes from the UV regime, i.e. from small values of $u$. On the other hand, the solution for $\Psi$ becomes unstable and noisy in this regime, so we usually choose a cutoff $u_{min}$ of the order of $10^{-5}$, which causes usually no significant errors in the result -- except for the case of the entropy. Even if we try to extrapolate at small $u$, the cancellation of the boundary term in \reef{entget} will not be accurate, so we need to push the minimum value for $u$ as far as possible and we will notice some noise. Eventually, the qualitative result will not be affected in either case, and we can further check whether some apparent ``effect'' is due to numerical errors or not by tuning $u_{min}$. 

We now use quantities that are made dimensionless using the mass as we consider the system at fixed mass. For example, we have $\bar{T} = \frac{1}{\tlm}$ and $\bar{B} = \frac{\tlb}{\tlm^2}$. We chose this combination, because these are parameters that naturally arise in the computations. Notice however, that $\tlm$ contains a factor of $\sqrt{\lambda}$, i.e. $\bar{T} =  \sqrt{\lambda} \frac{T}{2^{3/2} M_q} $ and $\bar{B} = \frac{\tlb}{\tlm^2} = \lambda \frac{B}{8 \pi^2 M_q^2} $.

\DFIGURE{\includegraphics[width = 0.49\textwidth]{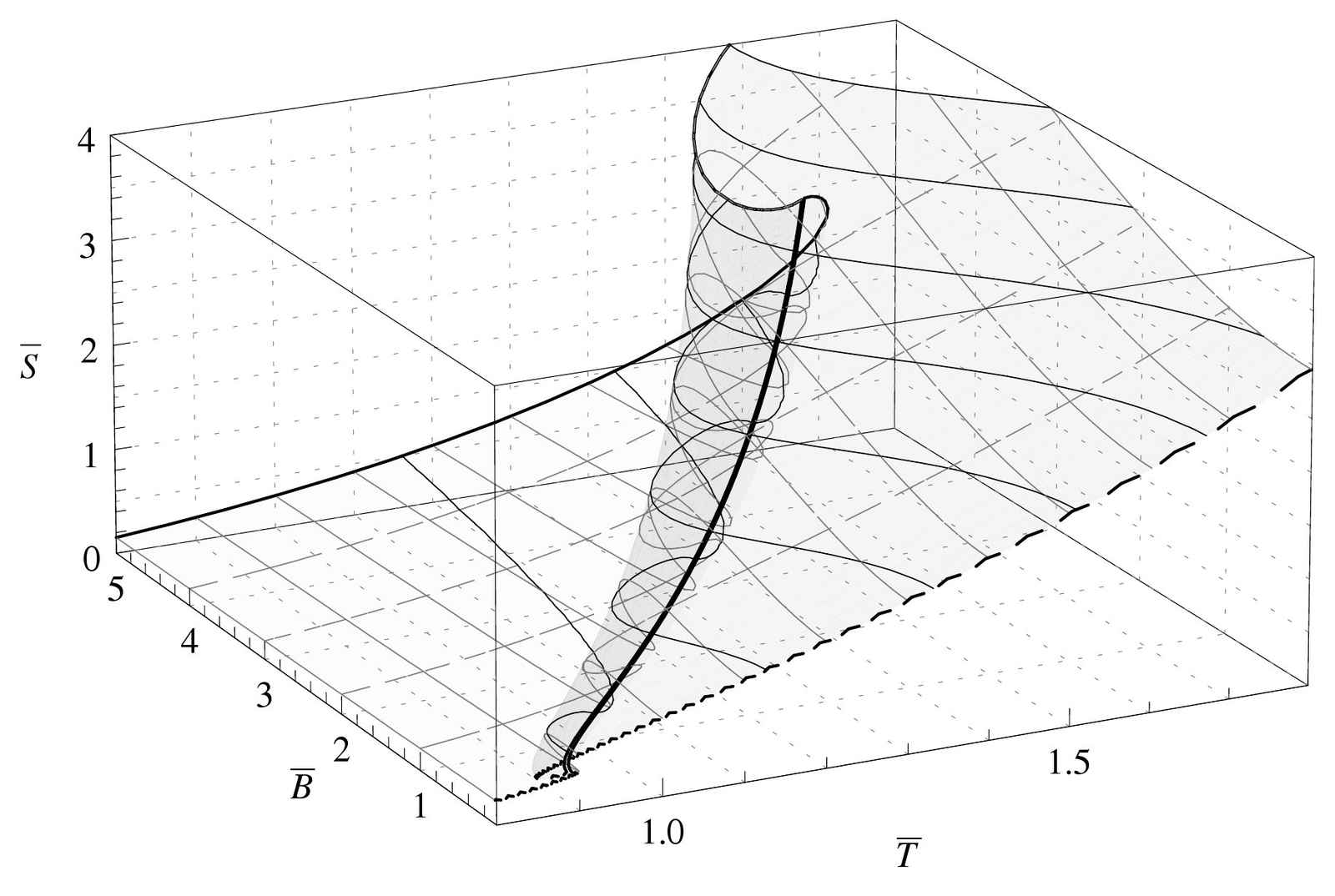}
\includegraphics[width = 0.49\textwidth]{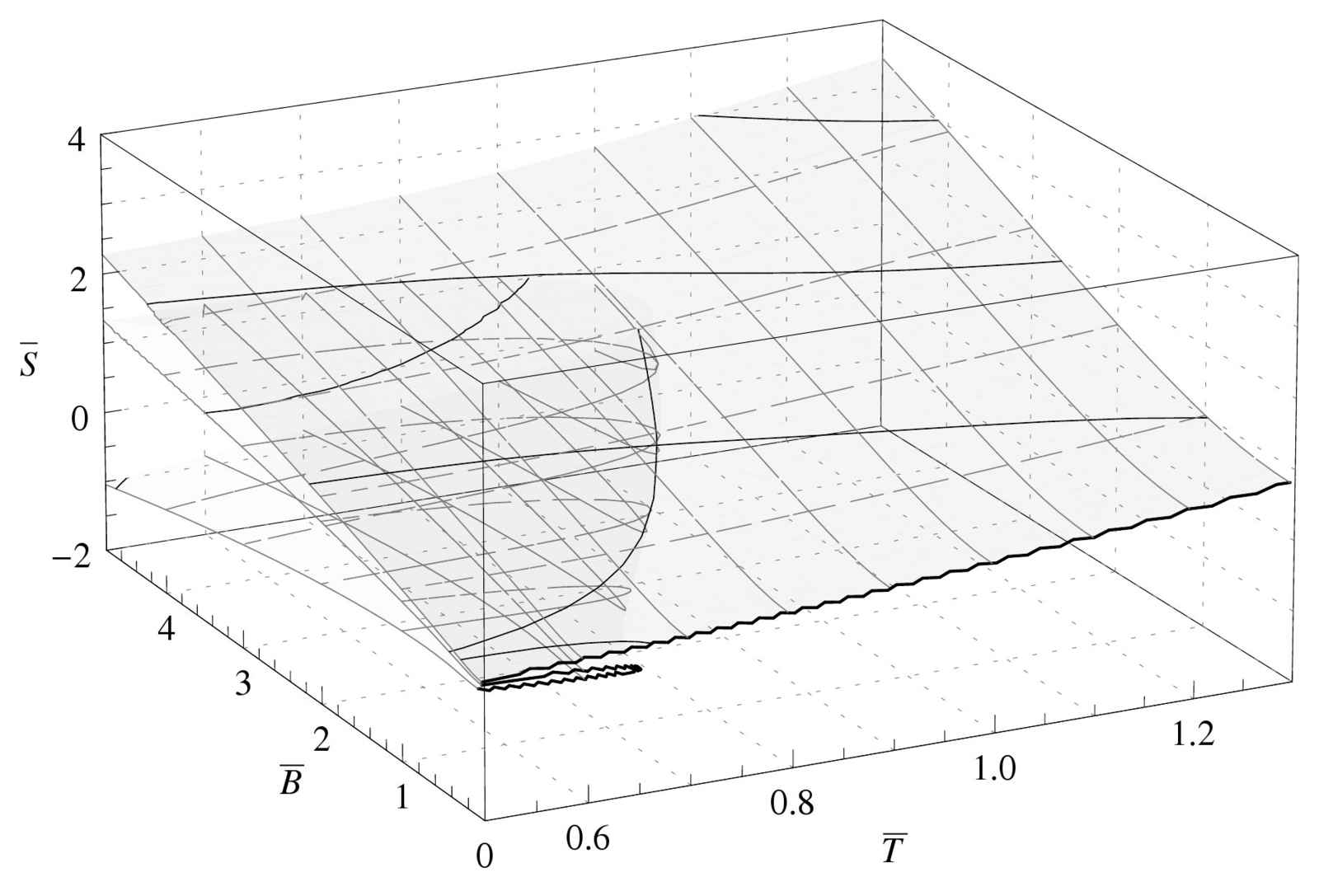}\\
\includegraphics[width = 0.49\textwidth]{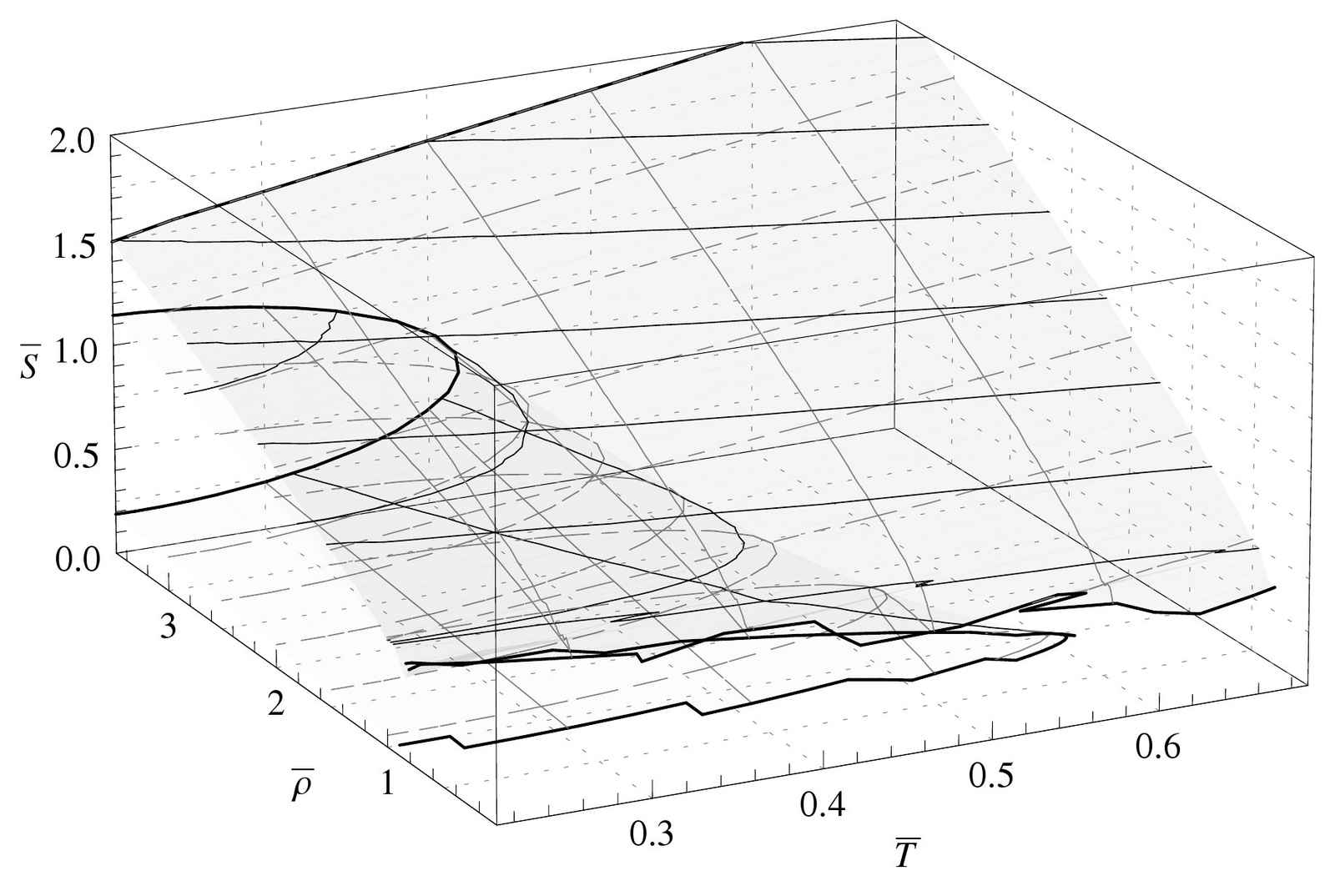}
\includegraphics[width = 0.49\textwidth]{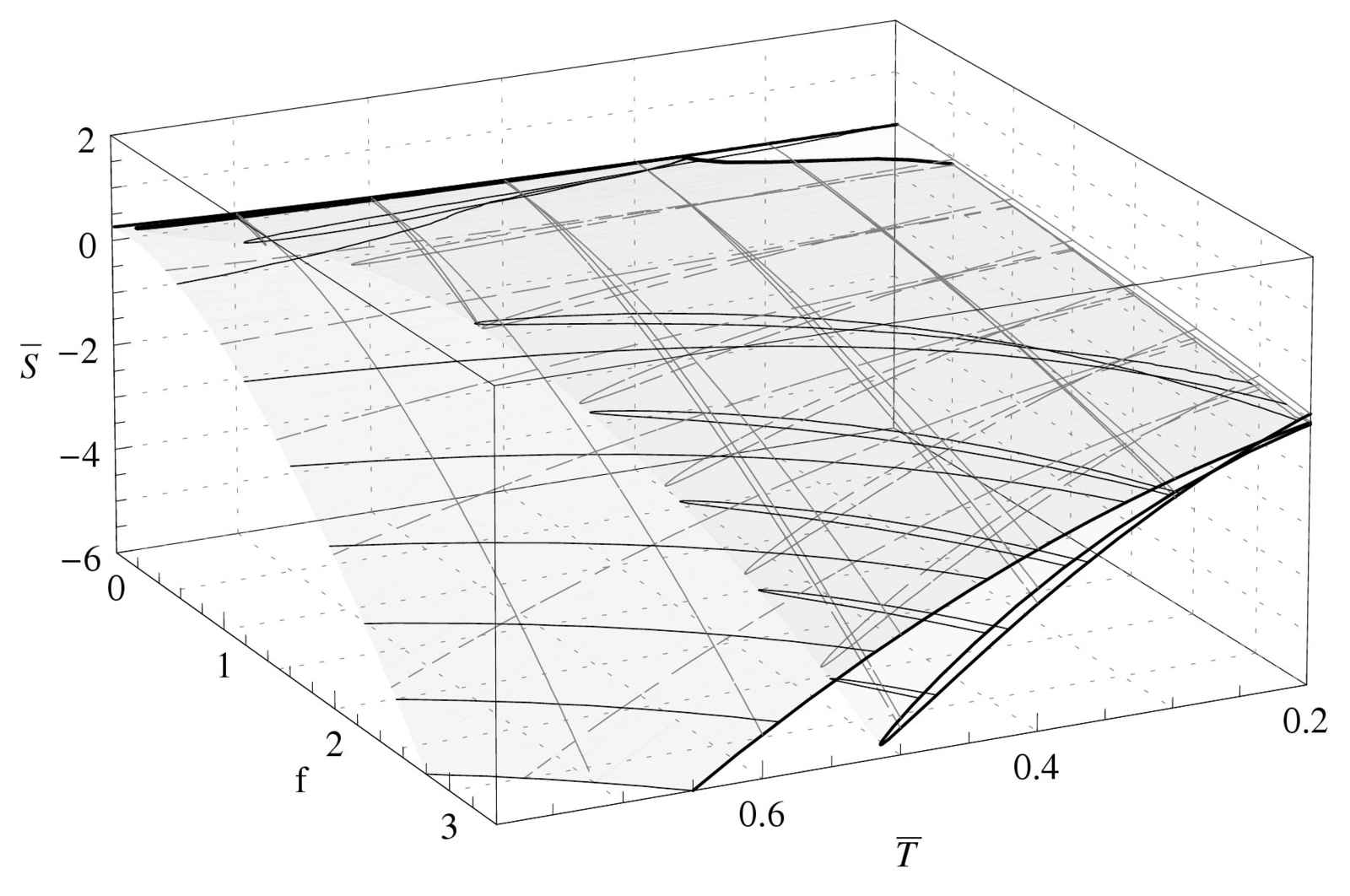}
\caption{The dimensionless entropy 
$\bar{S} =  \frac{S}{8 M_q^2 N_c N_f} $ as a function of $\bar{T} =  \sqrt{\lambda} \frac{T}{2^{3/2} M_q} $ and $\bar{B}= \lambda \frac{B}{8 \pi^2 M_q^2}$ for $f = 0 = \bar{\rho}$ (top left) and $f = 1/2$, $\bar{\rho}=0$ (top right) or as a function of $\bar{T}$ and $\bar{\rho}$ (bottom left) or $f$ (bottom right).}\mlabel{entroplot}}
In the results for the entropy in fig. \mref{entroplot}, let us first look at a few technical issues.  
In the plot on the top right, at $f=1/2$, we notice that there is some numerical error which causes the entropy of the lower branch of the Minkowski embedding to remain finite at vanishing temperature.  
In the figure on the bottom-left, we ``cut out'' the region of small temperatures and densities, as it was dominated by noise because the configuration in this region requires embeddings with $\Psi_0\sim 1$, which are numerically problematic.
Also, in  the plot in the bottom right, the hint of the entropy of the phases crossing around $\bar{T} \sim 0.25$ can be shown to be due to numerical errors. 

Looking at the results, we see that in all cases, the blackhole embedding is preferred. In the case of having only the magnetic field non-vanishing, we actually notice the cusp-like behavior at the point where the blackhole and Minkowski phases meet that was already observed for (3+1) dimensional systems in \cite{long}. We also notice again in the bottom right, which shows the plot ``from behind'', that $f$ causes a negative entropy contribution, as discussed in the context of fig. \mref{entropy_plot_M0}. 
The fact that the difference in entropy between the Minkowski and black hole phases vanishes as $T\rightarrow 0$ in the case of having the internal flux $f$ turned on, but that it doesn't vanish for the embeddings at finite density indicates that the latter are ``more unstable''. This reflects precisely the observation that Minkowski embeddings at finite density cannot be supported in string theory -- in contrast to Minkowski embeddings at finite flux $f$. From another point of view, the vanishing entropy at $T\rightarrow 0$ is what we expect generically and the finite entropy implies some ``degenerate ground state''.

\DFIGURE{\includegraphics[width = 0.49\textwidth]{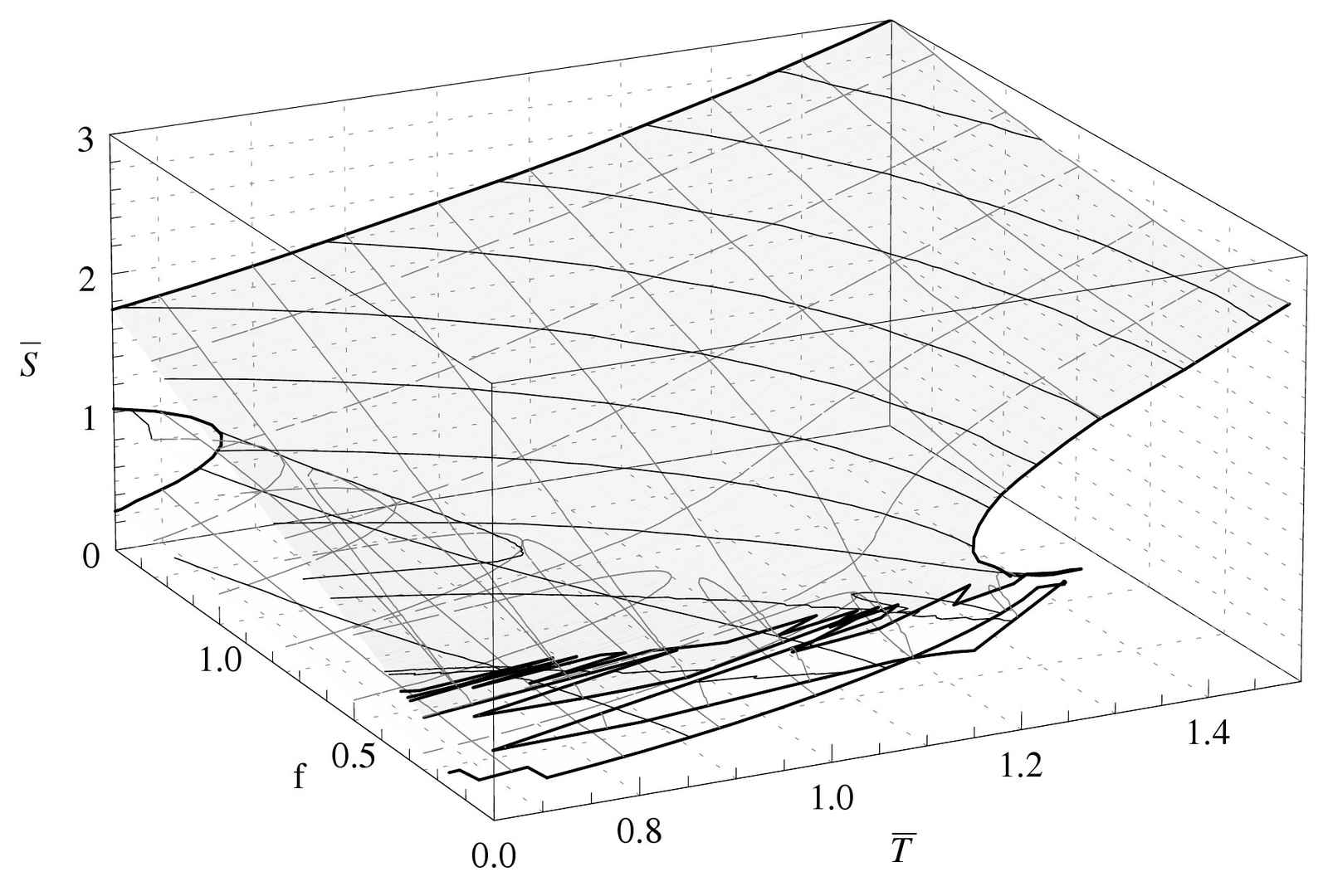}
\includegraphics[width = 0.49\textwidth]{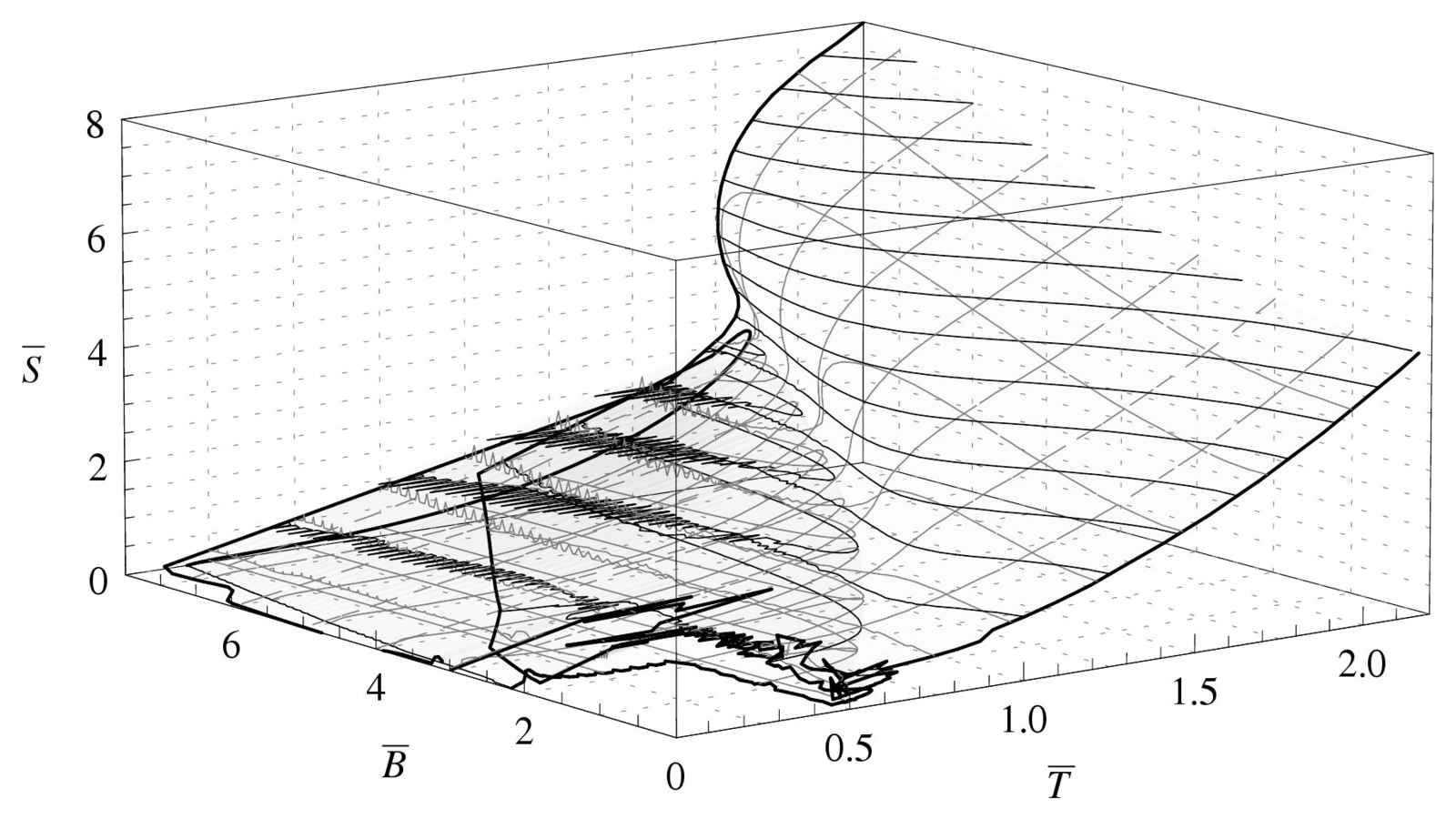}
\caption{Left: ``Zoom'' to the region of the phase transition between the different blackhole phase embeddings $\bf B1$ (low temperatures) and $\bf B$ (high temperatures) at small values of the density and fixed magnetic field $\bar{B}=2$. Right: the phase transition at large magnetic fields and fixed density $\bar{\rho}=0.5$}.\mlabel{phbarplot}}
Furthermore, we can see how the phase diagram changes if we consider a fixed quark mass, rather than a fixed temperature. $\tlm \rightarrow 0$ gets mapped to $\bar{T}\rightarrow \infty$ and is hence not accessible anymore, and in fact finite values of $\tlb$ also get mapped to $\bar{B} \rightarrow \infty$ in that limit. Hence, at all finite values of $\bar{B}$, there are both blackhole and Minkowski embeddings and the critical temperature $\bar{T}_c$ of the phase transition is some monotonously increasing function of $\bar{B}$. Since there is no indication of a transition between the phases {\bf B} and {\bf B1} in the plots in figure \mref{entroplot}, let us see why. In our new variables a trajectory of fixed $\bar{B}$ and varying $\bar{T}$ corresponds to a trajectory $\tlb = \frac{\bar{B}}{\bar{T}^2}$ in fig. \mref{phases_B_plot}. Hence we need to look at e.g. large values of $\bar{B}$ and $\bar{T}$ at fixed $\bar{\rho}$ or $f$ and the critical temperature will be $\bar{T}_c \propto \bar{B}^{1/2}$. If we expand the equation of motion for $\Psi$ around small values of $\Psi$ and study the resulting linear second order equation, we find that this phase transition always exists for sufficiently large $\bar{B}$ and $\bar{T}$. For finite values of $f\gtrsim 1$, the onset of the phase transition gets shifted to $\bar{B},\bar{T} \gg 1$, whereas for finite $\bar{\rho}\gtrsim 1$, the temperature scaling of $\bar{\rho}$ implies that the phase transition always appears at finite values of the temperature.
In fig. \ref{phbarplot}, we show how the phase transition appears at small values of $\bar{\rho}$ and fixed $\bar{B} = 2$, and at increasing values of $\bar{B}$ for fixed $\bar{\rho} = 0.5$.

In the light of the fact that the blackhole embedding is always thermodynamically preferred, the phase {\bf B1} and the phase transition between {\bf B} and {\bf B1} are a smooth continuation of the Minkowski phase and the blackhole-Minkowski embedding phase transition as we we turn on the quark density $\rho$. This is also reflected in the fact that the {\bf B1} phase has lower entropy due to the larger fraction of quarks bound in mesons.

\DFIGURE{\includegraphics[width = 0.49\textwidth]{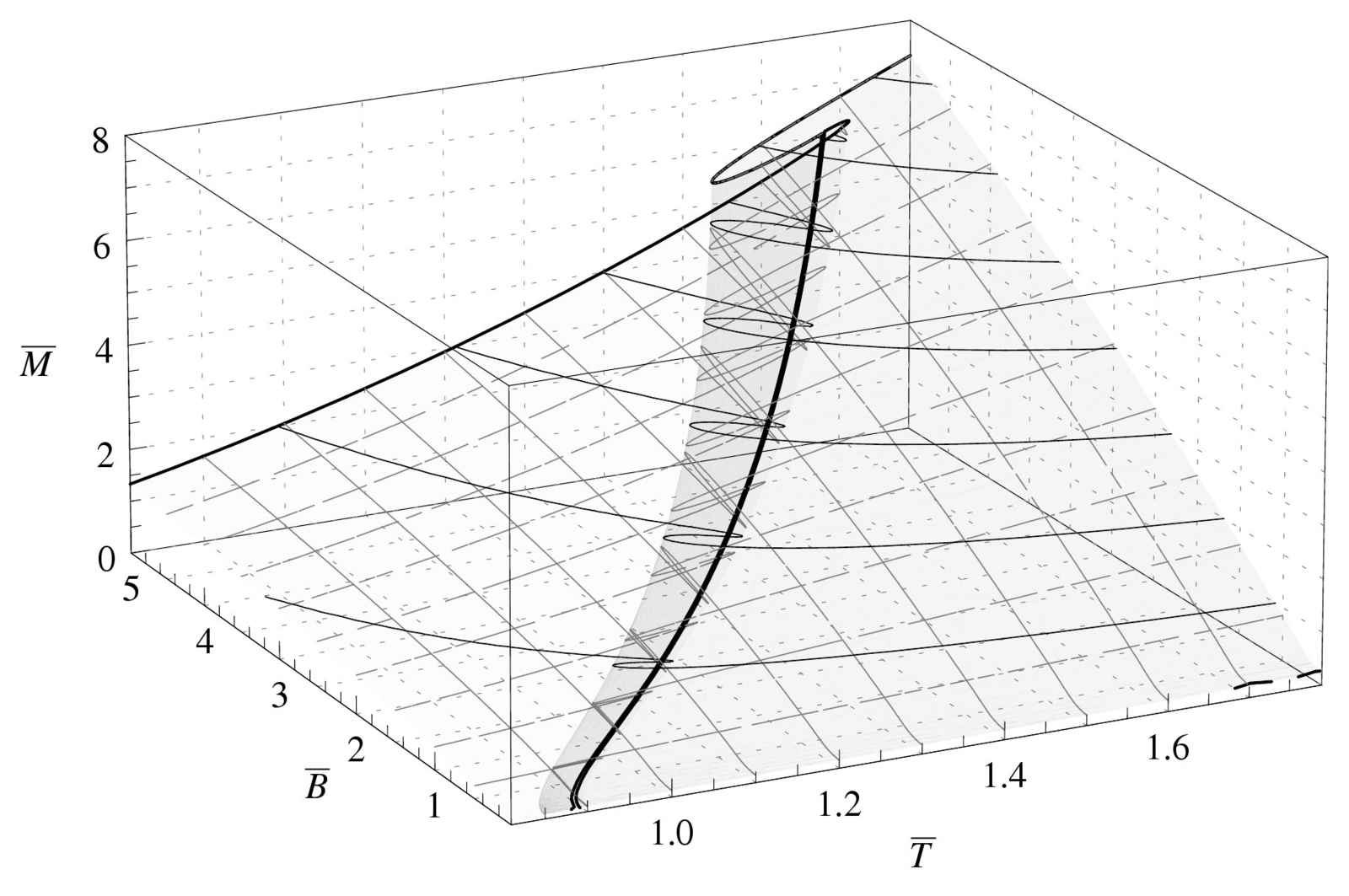}
\includegraphics[width = 0.49\textwidth]{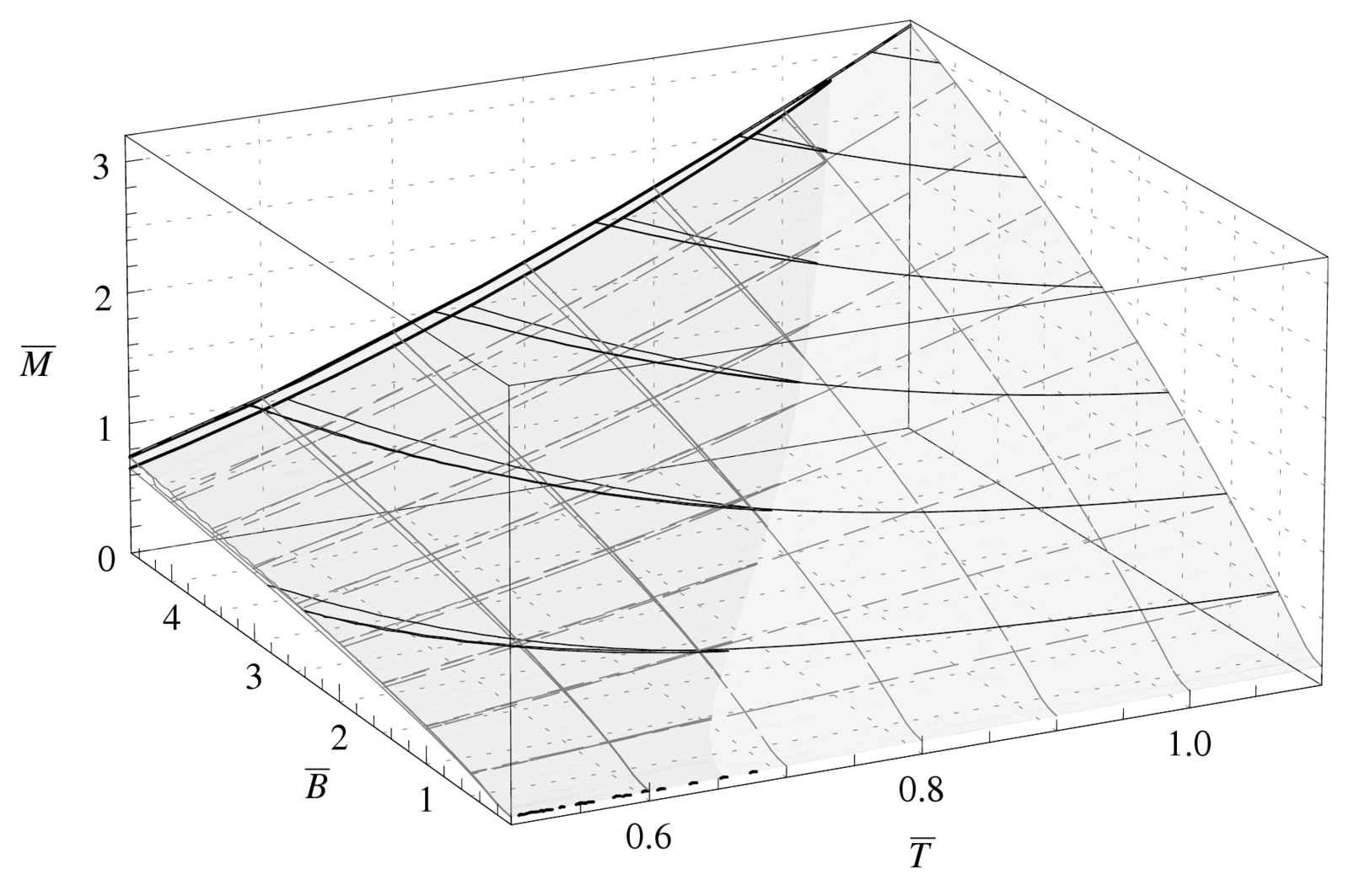}
\caption{The magnetization $\bar{M}$ as a function of the temperature-mass ratio $\bar{T}$ and the magnetic field $\bar{B}$ at $\bar{\rho} = 0 = f$ (left) and $\tlrho = 0$, $f=0.5$ (right).}\mlabel{magplot}}
Let us finally just take a quick look at the other derivatives of the free energy. 
In fig. \mref{magplot}, we show the magnetization and we notice that the magnetization is higher in the blackhole phase, which is dominated by free quarks, and lower in the Minkowski phase that are dominated mesons. In the case of small finite $f$ (qualitatively the same happens also for $\bar{\rho}$) however, the difference is highly suppressed and both phases have essentially the same magnetization.

\DFIGURE{\includegraphics[width = 0.49\textwidth]{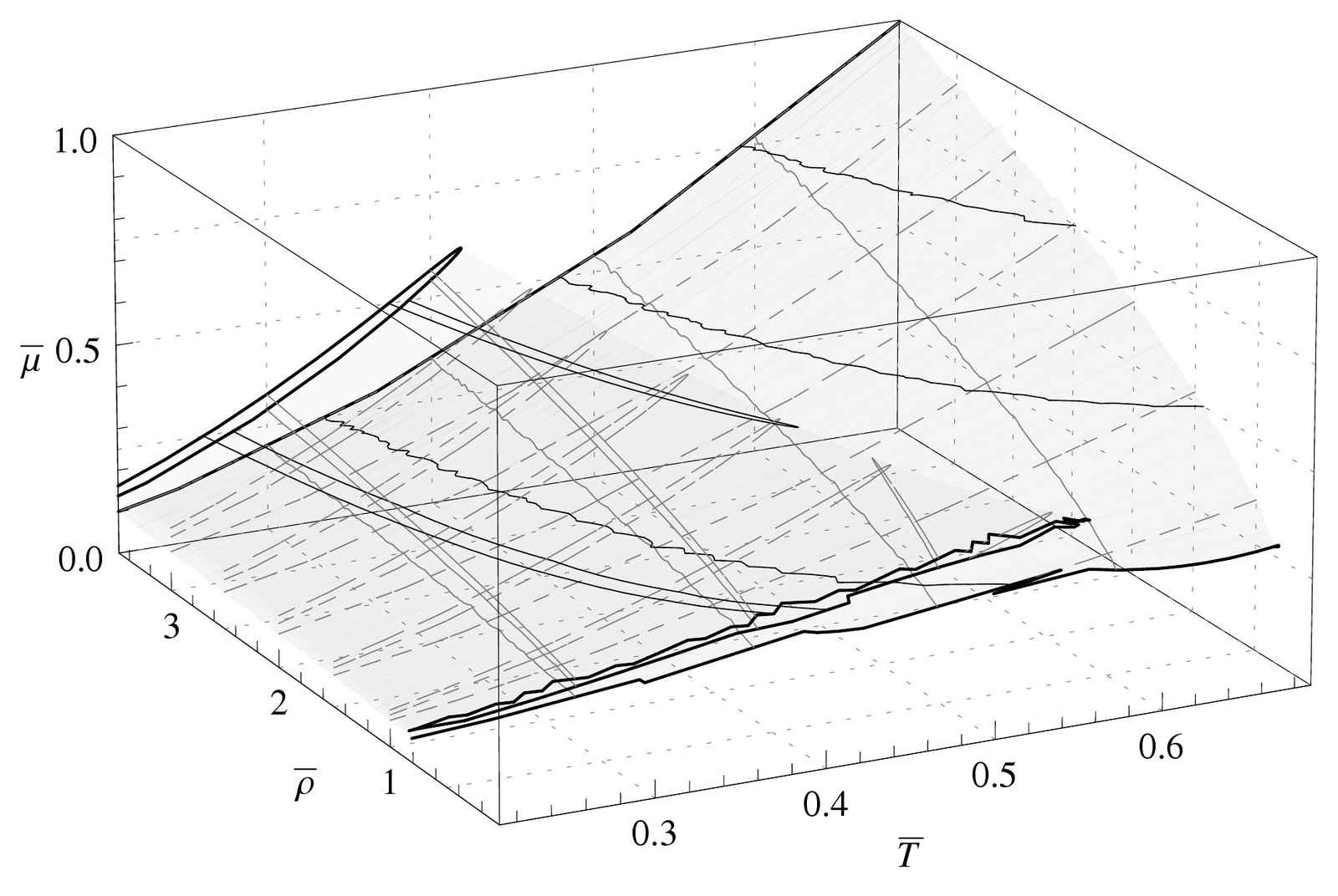}
\includegraphics[width = 0.49\textwidth]{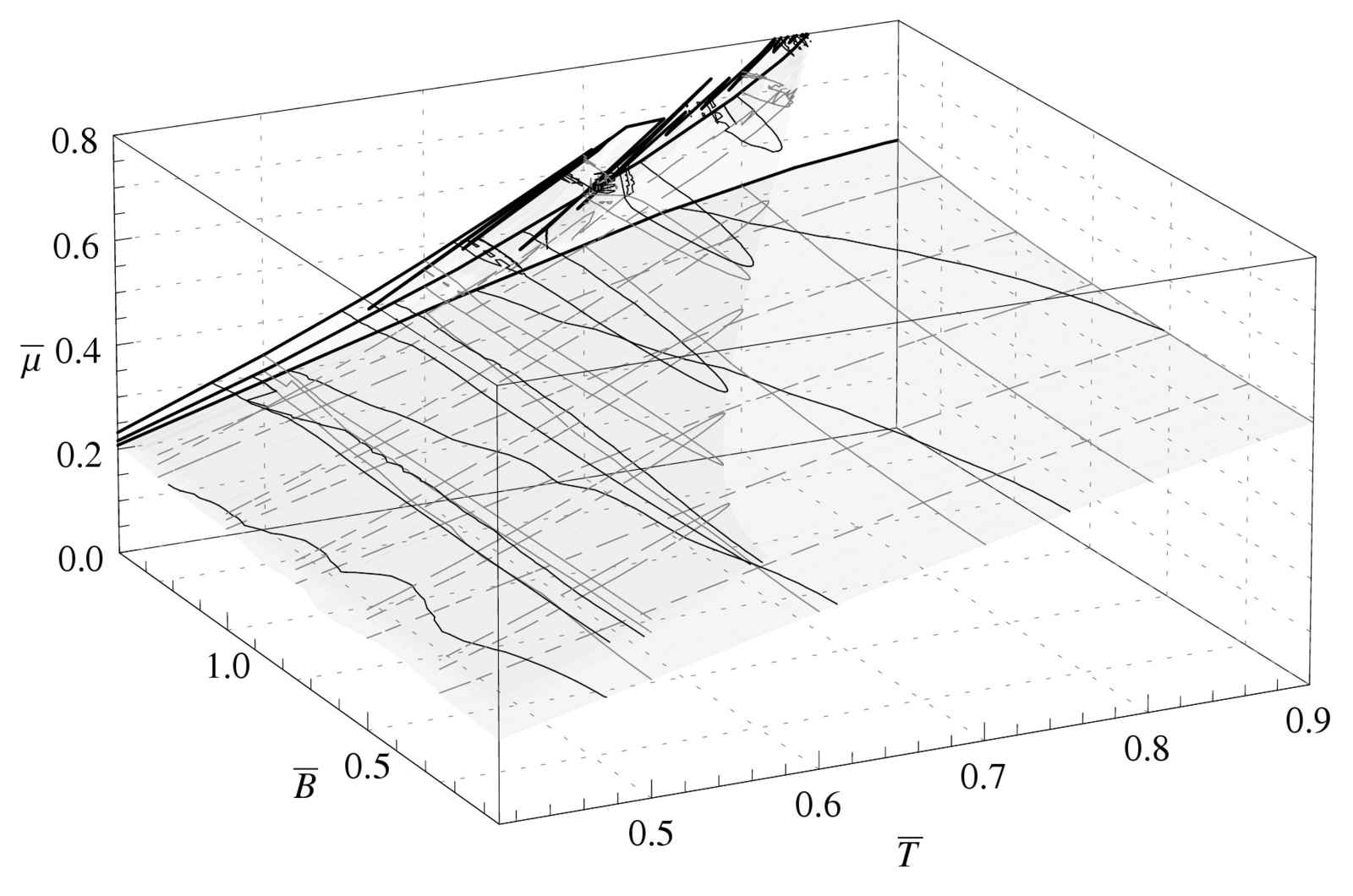}
\caption{The chemical potential $\bar{\mu}$ as a function of the temperature-mass ratio $\bar{T}$ and the density $\bar{\rho}$ at $\bar{B} = 0 = f$ (left) and as a function of $\bar{B}$ at $\bar{\rho}=1/2$ (right).}\mlabel{chemplot}}
In figure \mref{chemplot}, we look at the chemical potential. In contrast to the magnetization, there is a a significant difference in the chemical potential, that persists in all cases, with the one of the Minkowski phase being higher than the one of the black hole phase. Hence, inducing a difference in the density of quarks and anti-quarks requires more energy in the Minkowski phase. This arises intuitively, as the Minkowski phase is dominated by mesons. It might also be related to the fact that in string theory, there should be no finite net baryon density in the Minkowski phase.

\DFIGURE{\includegraphics[width = 0.49\textwidth]{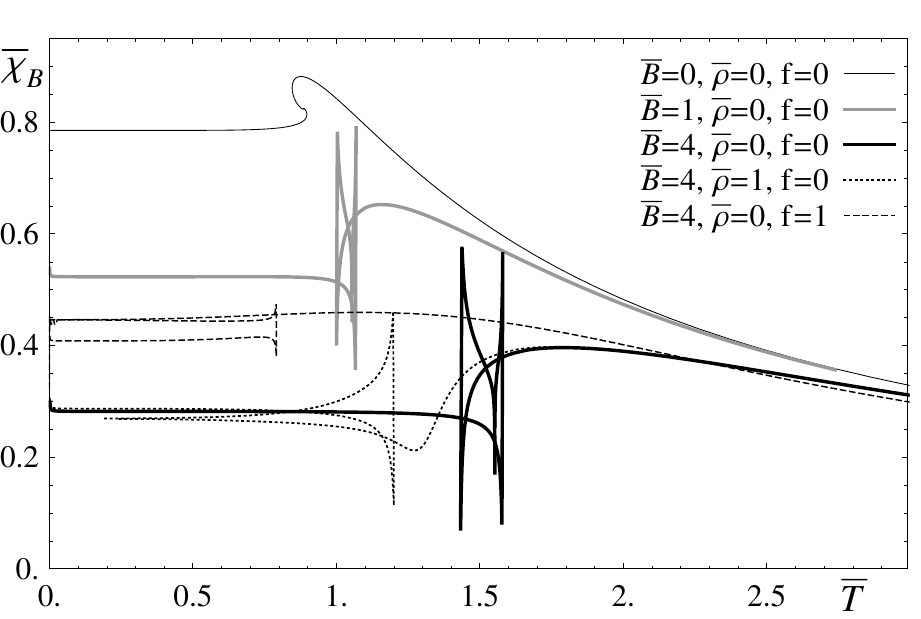}
\includegraphics[width = 0.49\textwidth]{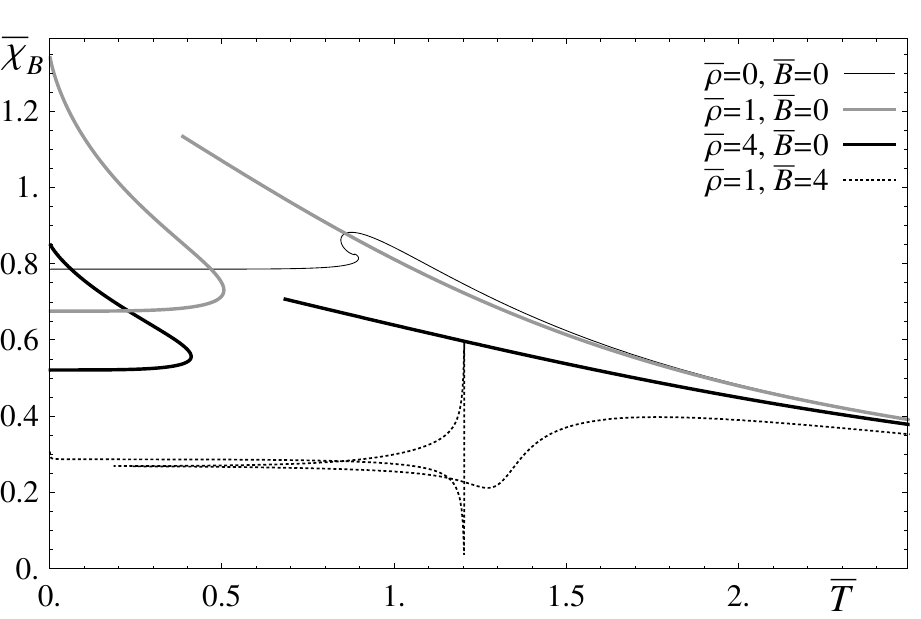}\\
\includegraphics[width = 0.49\textwidth]{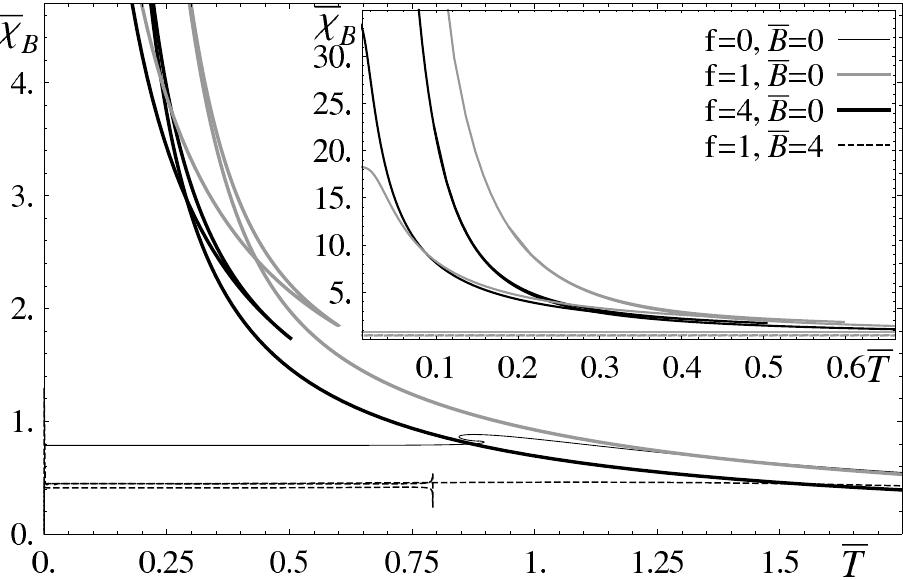}
\begin{minipage}[b]{0.49\textwidth}\caption{Dimensionless magnetic susceptibility of the defect, $\bar{\chi}_B = \frac{4\pi^2 \sqrt{2} M_q}{\lambda N_c N_f} \chi_B $ at finite magnetic field (top left), finite density (top right) and finite $f$ (left). The case $\bar{M}=\bar{\rho}=f=0$ is shown in all plots as a reference and divergent ``spikes'' are cut off at finite values for clarity.}\end{minipage}\mlabel{susM}} 
In the remainder of this section, we look at the second derivatives of the free energy that we obtained with the methods described in section \mref{responses}. 
In figure \mref{susM}, we show the magnetic susceptibility in the presence of the various parameters. In each of the plots, as also in figures \mref{densM} and \mref{heatM}, we show also the $f = \bar{\rho} = \bar{B} = 0$ case for reference. In the top left, at finite magnetic field, we observe the saturation behavior with increasing magnetic field and the phase transition with the temperature independent behavior in the Minkowski phase and the ``dimensional'' $\bar{T}^{-1}$ scaling in the blackhole phase. This is because the Minkowski phase is dominated by the physics and energies of mesons, whereas the blackhole phase is dominated by the statistical mechanics of free quarks in the plasma. Also, turning on a small density in the presence of the magnetic field changes the susceptibility only sightly, in particular around the region of the phase transition, but turning a finite $f$ changes it more significantly. Hence, apparently the net quark density is less relevant than the total quark density.
Also the large temperature ``tail'' is universal, as it is dominated by the thermal equilibrium of ``quark-anti-quark'' production. At vanishing magnetic field, we see no divergent susceptibility at the phase transition, which is expected as there is no remnant magnetization at $B=0$ and hence also the ``latent magnetization'' of the phase transition vanishes. 
In the presence of (only) a finite density, it appears that the high temperature blackhole phase also extends to small temperatures. For numerical reasons as we will discuss in the below, we cannot trace this phase to vanishing temperatures. However, the one of the fiducial Minkowski phases that typically mimics the blackhole phase is finite at vanishing temperature. As we saw in the massless case in section \mref{massless}, increasing the density lowers the magnetic response.
In the presence of (only) finite $f$, the behavior is significantly different. On the one hand, the curves do not converge in the high temperature regime, mainly because of the absence of a temperature scaling of the parameter $f$. At large temperatures, we still have the effect that finite $f$ lowers the susceptibility, but at small temperatures, the susceptibility increases. Tracking the ``lower'' Minkowski phase to $T\rightarrow 0$ shows actually that the lines of $f=1$ and $f=4$ cross and the zero temperature magnetization remains finite. 
The (stable and hence more relevant) blackhole phase however appears to diverge at small temperatures -- but the numerics start to fail in this branch around $T\sim 0.05$ ($\Psi_0 \sim 1- 5 \times 10^{-4}$). This means that the defect is becoming magnetized. However, our methods are not sufficient to explore whether there is some spontaneous magnetization at vanishing temperature.

\DFIGURE{\includegraphics[width = 0.49\textwidth]{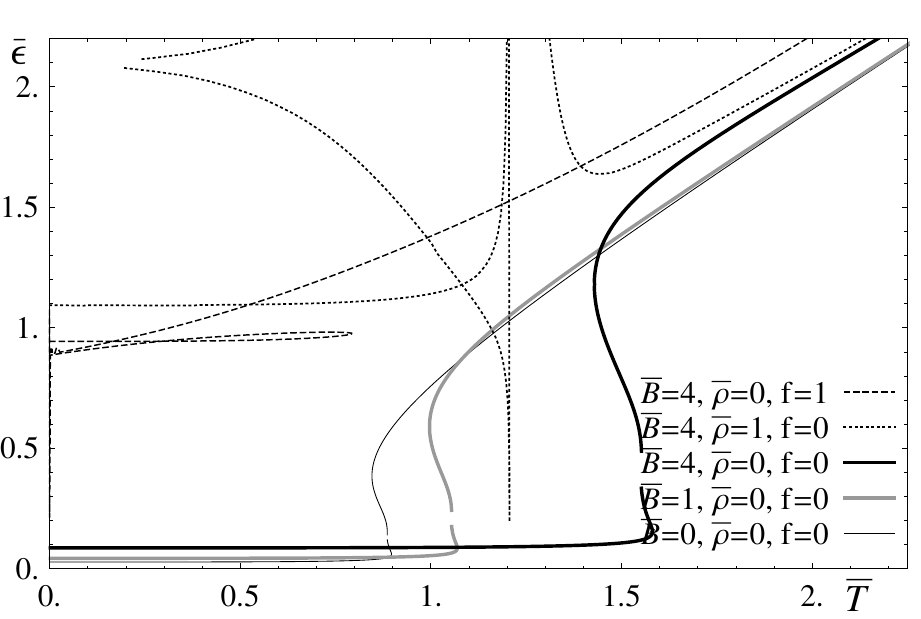}
\includegraphics[width = 0.49\textwidth]{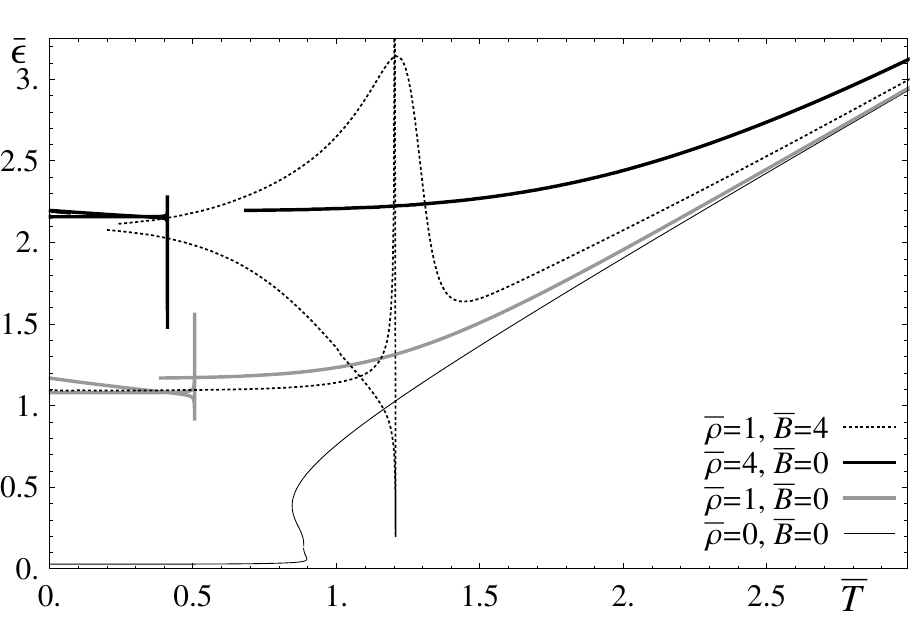}\\
\includegraphics[width = 0.49\textwidth]{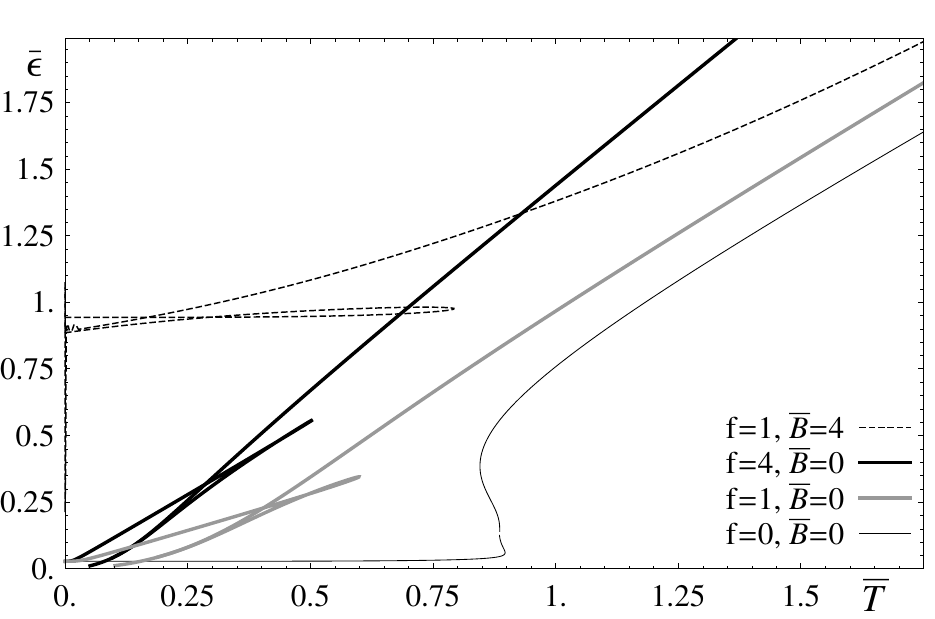}
\begin{minipage}[b]{0.49\textwidth}\caption{Dimensionless density of states of the defect, $\bar{\epsilon} = \frac{4\pi^2 \sqrt{2}}{M_q N_c N_f} \epsilon$ at finite magnetic field (top left), finite density (top right) and finite $f$ (left). The case $\bar{M}=\bar{\rho}=f=0$ is shown in all plots as a reference and divergent ``spikes'' are cut off at finite values for clarity.}\end{minipage}\mlabel{densM}}
The ``density of states'', i.e the inverse of the response of the chemical potential to the net quark density, is shown in figure \mref{densM}.
In the Minkowski phase, $\epsilon$ is highly suppressed at some small, but constant, value as we expect for the mesonic phase. At finite magnetic fields, this increases with some multiplicative factor.
This just means that the chemical potential decreases with increasing magnetic field, and there is some rapid increase as we turn on a small density.
At large temperatures, however, all curves converge to the $f = \bar{\rho} = \bar{B} = 0$ case. 
As we turn on a finite density, this suppressed density of states is lifted to larger, but still approximately constant values, now in the blackhole phase that ``allows'' free quarks. 
That is the case if inducing a net quark density corresponds to increasing the total quark density -- with a finite quark mass. If we turn on a magnetic field in addition to the finite density, there is a strong increase in $\epsilon$ in the low temperature blackhole phase. A wild speculation would be that this is related to the density of Landau levels. Just as in the case of the susceptibility, the high-temperature behavior depends significantly on $f$. This may be simply because increasing $f$ increases the physical ``width'' of the defect. As it is the case with the other quantities, there is no apparent phase transition in the blackhole phase. Now it seems, that in the blackhole phase $\epsilon$ vanishes as $T\rightarrow 0$, which is similar to the diverging magnetic susceptibility. In the presence of a small magnetic field, surprisingly, the low temperature behavior does not get scaled by some factor as in the case of $f=0$, but it actually gets substantially shifted to some larger value.

\DFIGURE{\includegraphics[width = 0.49\textwidth]{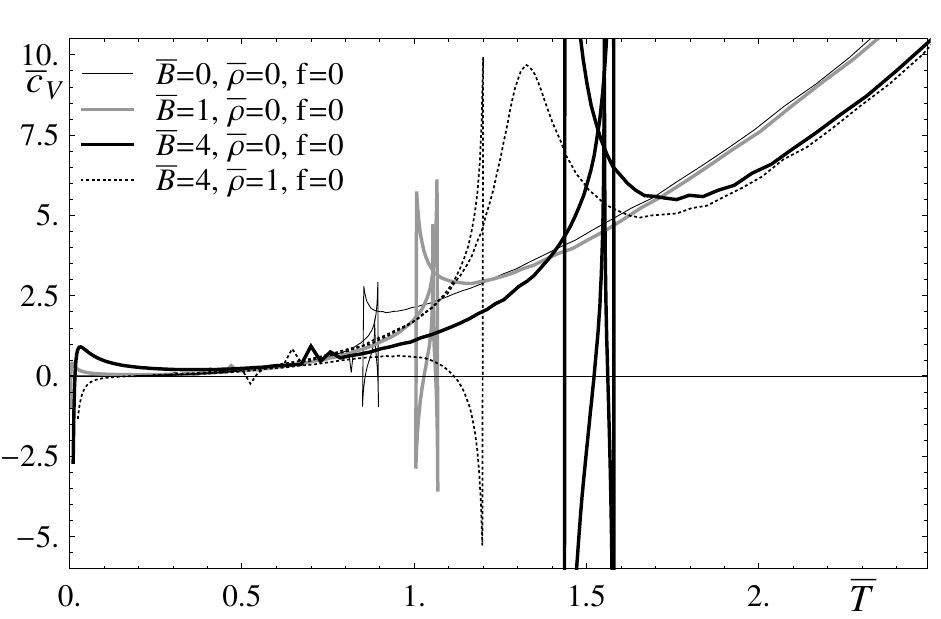}
\includegraphics[width = 0.49\textwidth]{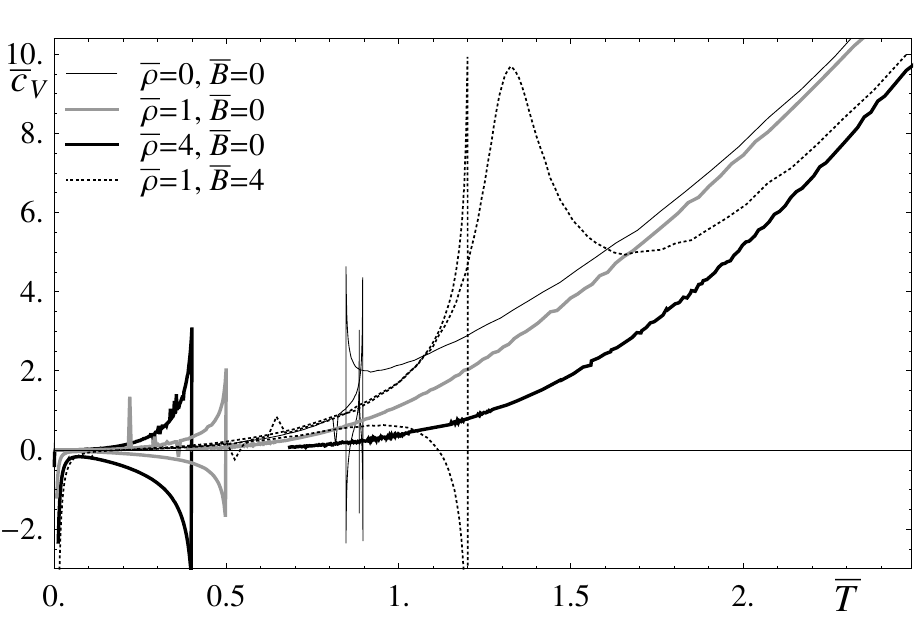}\\
\includegraphics[width = 0.49\textwidth]{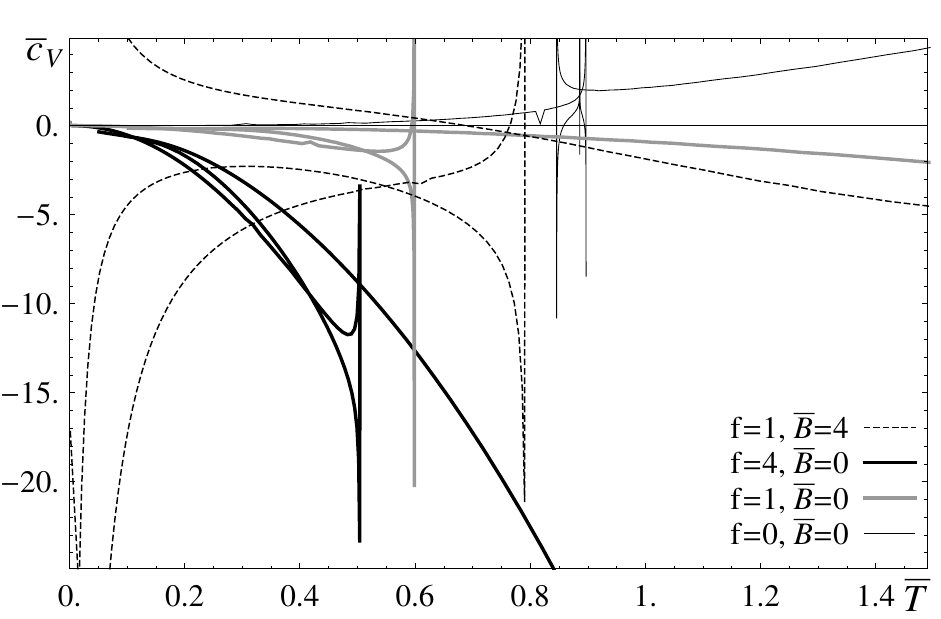}
\begin{minipage}[b]{0.49\textwidth}\caption{Dimensionless heat capacity of the defect, $\bar{c}_V = \frac{ \sqrt{\lambda}}{8 M_q^2 N_c N_f} c_V$ at finite magnetic field (top left), finite density (top right) and finite $f$ (left). The case $\bar{M}=\bar{\rho}=f=0$ is shown in all plots as a reference and divergent ``spikes'' are cut off at finite values for clarity.}\end{minipage}\mlabel{heatM}}
Finally, we study the results for the heat capacity in figure \mref{heatM}. There, we see that there is no significant difference between the Minkowski and blackhole phases, except for the diverging heat capacity around the phase transition and some shift between the phases in the $f = \bar{\rho} = \bar{B} = 0$ case with a higher value in the blackhole phase, because of the extra degrees of freedom of free quarks. Increasing the magnetic field slightly lowers the heat capacity, as does turning on a finite density. We can also nicely observe the peak in the heat capacity around the crossover between the high and low temperature blackhole phases at $\bar{\rho}=1$, $\bar{B}=4$. There are a divergence at $T=0$ and some noise at small temperatures, both of which are due to numerical inaccuracy. There are two more noteworthy observations. One is the negative heat capacity at finite $f$. This effect is simply due to the fact that the extra degrees of freedom from modifying $N_c \rightarrow N_c + \delta N_c$ are not available from $z=0$ but turn on only gradually over the defect -- as we discussed in the context of the entropy in section \mref{massless}. The numerics at $f=1$, $\bar{B}=4$ at small temperatures seem to be not reliable.

The other interesting observation is the fact that the heat capacity vanishes very quickly with decreasing temperatures in the blackhole phase at finite densities. As mentioned before, the numerics let us down in this case at $\bar{T}\sim 0.7$ in the $\bar{\rho} = 4$ case and around $\bar{T}\sim 0.74$ for $\bar{\rho} = 1$, but already at these values $c_v$ has converged to $0$ much faster than expected for some power law scaling. This region is very interesting, as it turns out that at fixed $\bar{\rho} = \frac{\tlrho}{\tlm^2}$, large values of $\tlm = \frac{1}{\bar{T}}$ are obtained at some ``finite'' $\Psi_0 \in ]0,1[$ and not at $\Psi_0\rightarrow 1$. This arises because at large $\tlm$ and $\tlrho$, we have $\tlm(\Psi_0, \tlrho) \propto \tlrho^2$. The value of $\Psi_0$ determines the proportionality factor, such that we get a value of $\Psi_0$ depending on $\bar{\rho}$. Such coincidence in the scaling arises generically only in $d_{def.}+1$ dimensional matter embedded in $d_{bulk} + 1 = 2 d_{def.}$ dimensions. From a physical point of view in the gravity picture, the probe brane becomes ``stiffer'' with increasing $\tlrho$, which causes some ``straight'' limiting profile as we take $\bar{T} \rightarrow 0$.  This limit can however not be suitably studied with the methods of this paper, as we tune $\Psi_0$ (or $u_{max}$) to obtain $\tlm$. A similar behavior also occurs in the corresponding branch of the (fiducial) Minkowski embedding.
%
%
%
%
%
\section{Discussion and Conclusions} \mlabel{discuss}
In this paper, we applied the holographic duality to study the thermodynamic properties of a class $(2+1)$-dimensional defect CFT's emersed in a heat bath of 3+1 dimensional $\mathcal{N} = 4$ $SU(N_c)$ SYM theory. To do that, we considered the properties that are extrinsic in the ``area'' of the defect, i.e. from a $(3+1)$-dimensional point of view are independent of the placement of boundaries at some $z_{bdy}$ away from the defect -- in the limit of large $z_{bdy}$.

We considered primarily the string theory setup realized by
embedding $\nf$ probe D5-branes in an AdS$_5\times S^5$ background, in which case the system (at $T=0$)
preserves eight supersymmetries. 
At vanishing quark mass, the thermodynamic properties are identical (modulo overall factors) to the defect CFT dual to embedding $\nf$ probe D7-branes in the AdS$_5\times S^5$, which preserves no supersymmetries. As this contains only fermions at the massless level, it is of particular interest to condensed matter physics. As discussed in \cite{fancycon} the results of these embeddings are not reliable because it is  not quite sure in how far the D7 setup persists in the light of gravitational backreaction. 

As we outlined in section \mref{gravsetup}, we considered deformations to the theory corresponding to a shift $N_c \rightarrow N_c + \delta N_c$ in the level of the gauge group on one side of the defect parametrized by $f = \frac{\pi \,\delta N_c}{\sqrt{\lambda} N_f} $, to a finite quark mass $M_q$, finite net baryon number density $\rho$ and magnetic field $B$. 

First, in section \mref{thermdefs}, we discussed the choice of thermodynamic variables. We Legendre-transformed the action in order to use the density as a thermodynamic variable, rather than the chemical potential. There we also demonstrated that the parameter $f$ is represented by the change in (3+1) free energy and its dual, defect width. As $f$ also parametrizes the shift in the level of the gauge group $\delta N_c$ we chose to keep it as a fixed parameter, for which we also needed to Legendre transform with respect to $\partial_u z$. In that context, we demonstrated how the tension of the probe brane at its endpoint in the case of Minkowski embeddings is matched by the appropriate number of attached D3 branes.

We then derived the response functions, which allowed us to verify the expressions for the chemical potential and dual condensate assumed in section \mref{fancyintro}. We also developed a method to obtain the second derivative of the thermodynamic potential in closed-form, i.e. without taking numerical derivatives.

In the massless case, we were able to obtain analytic expressions for all the thermodynamic quantities. We presented those in section \mref{massless}, were we also discussed the negative entropy of the defect, and verified that it does not imply a negative entropy region and is simply related to the fact that the extra degrees of freedom due to $\delta N_c$ turn on only gradually over the ``width'' of the defect. We also observed the implications of the electromagnetic duality that was discussed in \cite{fancycon}. Here it corresponds again to an exchange of dimensionless density and magnetic field, $\tlrho \leftrightarrow \tlb$, and e.g. an exchange of chemical potential and magnetization. Overall, we found a saturating behavior, where e.g. the chemical potential scales as $\tilde{\mu} \propto \sqrt{\tlrho}$ at large densities.

In section \mref{massive}, we discussed the massive case. Using the dual condensate as an order parameter, we mapped out the phase diagram. We found that $\tlc$ vanishes in the blackhole phase but not in the Minkowski phase, where it increases with increasing magnetic fields; and that the critical mass of the phase transition decreases, i.e. the critical temperature increases, with increasing magnetic field similar to what was observed in \cite{magnetic,Filev}. In the case of a finite density or finite $f$, the blackhole phase extends to large dimensionless masses, i.e. vanishing temperature and the Minkowski phase splits into two metastable phases that extend to large mass or vanishing temperature. This deformation is smooth, and at small values of $\tlrho$ and $f$ the blackhole phase tracks closely the Minkowski phase, and hence we find a transition between two different blackhole phases. At larger $\tlrho$, $f$, this transition disappears and also the two Minkowski phases split further -- one remains approximately unchanged and the other one approaches the blackhole phase. The phase transition, or minimum mass/largest temperature of the Minkowski phase shifts towards lager masses, approximately $\tlm_{crit.} \propto \sqrt{\tlrho},\sqrt{f}$.

We then studied the results for the response functions. To do so, we considered fixed mass, and quantities made dimensionless with the appropriate powers of the combination $2 \sqrt{2/\lambda}M_q$, rather than $\pi T$. First, we looked at the entropy and found that the blackhole phase has higher entropy than the Minkowski phase, as expected since it is dominated by free quarks rather than mesons in the Minkowski phase. The entropy vanishes at vanishing temperature within the limitations of numerical accuracy, except for the blackhole phase at finite density, where there is some ``ground state degeneracy''. We also demonstrated how the phase diagram gets mapped in this parametrization of fixed mass; e.g. the phase transition at large magnetic fields is at $T\propto \sqrt{B}$. 

Overall, in the high temperature regime, the effect of fixed $B$ and $\rho$ disappears, because the relevant parameters are the dimensionless $\tlb$ and $\tlrho$ that scale $\propto T^{-2}$, but the effect of fixed $f$ obviously remains. The behavior in this regime is verified to be consistent with the massless limit. 

At small temperatures, there are a few interesting effects. At finite $f$ the crossover between the blackhole phases disappears and one of the Minkowski phases is very similar to the blackhole phase. For the magnetic response, we find that at finite $f$, the defect seems to become magnetized at vanishing temperature, as the susceptibility diverges at small temperatures and vanishing magnetic field but is finite and constant at finite magnetic fields. In the Minkowski phase, it becomes large, but finite. A similar behavior occurs with the response to the density, where the ``density of states'', the inverse of the second derivative of the free energy, vanishes. Certainly, it would be interesting to see whether there is a spontaneous symmetry breaking at vanishing temperature and finite $M_q$ and $f$, but this is beyond the scope of this paper.

At $\rho = B = f = 0$ and in the presence of finite magnetic fields, the density of states below the phase transition is small but finite and approximately temperature independent and increases with increasing magnetic field. The susceptibility is also approximately constant, but of $\order(1)$ and decreases consistent with the magnetic saturation effects. In the case of finite density, the density of states is also $\order(1)$ and approximately constant at low temperatures, increasing with increasing density consistent with saturation effects and the susceptibility is finite and increases with decreasing temperature in a smooth continuation of the high temperature behavior. 

A big surprise however happens for the heat capacity, which vanishes at finite $\rho$ with decreasing temperatures more quickly and at larger temperatures than expected for the general $T^2$ behavior. This is particularly interesting in the context of the unusual embedding in this regime. Rather than at $\Psi_0\rightarrow 1$, i.e. the limit that the probe brane intersects the horizon only at a point, we obtain $T\rightarrow 0$ at fixed $M_q$ and $\rho$ at some smaller value of $\Psi_0$. This arises because of a coincidence in the dimensionless scaling of the quark mass $M_q/T \propto \sqrt{\rho/T^2}$ at small temperatures, with a proportionality constant depending on $\Psi_0$ and hence relating $\rho/M_q^2$ to $\Psi_0$. Physically, this means that the brane has a higher effective tension, i.e. becomes ``stiffer'' as we decrease the temperature at fixed $\rho$, i.e. as we increase $\rho/T^2$ in such a way that there is a limiting ``straight'' embedding at fixed $\Psi_0$ as we decrease the temperature at fixed mass, i.e. as we increase $M_q/T$.
It would be very interesting to explore this new behavior further, but we will postpone this to future research as the methods used for this paper are not suitable and break down at small, but finite temperatures in this regime.

\acknowledgments 
I would like to thank Rob Myers for many helpful discussions and suggestions and for proofreading an earlier draft of this paper and
Yun-Seok Seo and Sang-Jin Sin
for helpful discussions and useful comments.
Research at the Perimeter Institute
is supported in part by the Government of Canada through NSERC and
by the Province of Ontario through MRI. This research is also supported
from an NSERC Discovery grant, from the Canadian Institute for
Advanced Research and by the Korea Science and Engineering Foundation (KOSEF) grant funded by the Korea Government (MEST) through the Center for Quantum Spacetime (CQUeST) of Sogang University with grant number R11-2005-021.
%
%
%
%
%
%
%

\end{document}